\documentclass[twocolumn]{aa}
\usepackage{graphicx}
\usepackage{natbib}

\newcommand{\dg}{$^\circ$}
\newcommand{\COt}{\mbox{CO($J$=2--1)}}
\newcommand{\COo}{\mbox{CO($J$=1--0)}}

\newcommand{\hmm}{{\rm H}_2}            

\newcommand{\ico}{I_{\rm CO}}

\newcommand{\XCO}{$X_{\rm CO}$}         

\newcommand{\nine}{\times 10^9~{\rm M}_\odot}            
\newcommand{\Tmb}{T_{\rm mb}}
\begin{document}

   \title{Molecular gas in the galaxy M\,83}
   \subtitle{I. The molecular gas distribution}

\author{A.~A.~Lundgren\inst{1,2} \and T.~Wiklind\inst{3,4}
\and H.~Olofsson\inst{1} \and G.~Rydbeck\inst{4}}

\institute{Stockholm Observatory, AlbaNova, SE-106 91 Stockholm, Sweden
\and European Southern Observatory, Casilla 19001, Santiago 19, Chile
\and Space Telescope Science Institute, 3700 San Martin Drive Baltimore, MD 21218, USA
\and Onsala Space Observatory, SE-43992 Onsala, Sweden}

\offprints{Andreas Andersson Lundgren (andreas@astro.su.se)}

\date{Received $<$date$>$ / Accepted $<$date$>$}

\abstract{

We present $^{12}$CO $J$=1--0 and $J$=2--1 Swedish-ESO Submillimetre
Telescope (SEST) observations of the
barred spiral galaxy M83 (NGC5236). The size of the CO maps is
10\arcmin$\times$10\arcmin\ and they cover the entire optical disk.
The grid spacing is 11\arcsec\ for \COo\ and 11\arcsec\ or
7\arcsec\ for \COt\ depending on the position in the galaxy. In total
we have obtained spectra in 1900 and 2574 positions in the \COo\ and
\COt\ lines, respectively. The CO emission is strongly peaked toward the
nucleus, which breaks up into two separate components in the \COt\ data
due to the higher spatial resolution.  Emission from the bar is 
strong, in particular on the leading edges of the bar.  The molecular
gas arms are clearly resolved and can be traced for more than 360\degr .
Emission in the inter-arm regions is detected.  The average CO
(\mbox{$J$=2--1})/(\mbox{$J$=1--0}) line ratio is 0.77. 
The ratio is lower than this on the spiral arms and higher in the 
inter-arm regions.
The arms show regularly spaced concentrations of molecular gas, 
Giant Molecular Associations (GMA's), whose masses are of the order 
10$^7$\,M$_\odot$. 
The total molecular gas mass is
estimated to be 3.9$\times$10$^9$\,M$_\odot$.  This mass is comparable
to the total HI mass, but H$_2$ dominates in the optical disk.  In the
disk, H$_2$ and HI show very similar distributions, including small
scale clumping.
We compare the molecular gas distribution with those of other
star formation tracers, such as B and H$\alpha$ images.

 }

\maketitle


\section{Introduction} In disk galaxies the mass of the molecular (H$_2$) gas is usually
similar to, or a bit less than, that of the atomic (HI) gas
\citep{YS91,CSG98}.  However, their spatial distributions are
completely different, both in the radial and in the vertical
directions.  The molecular gas is concentrated to a relatively thin
disk, which often has a radial extent similar to that of the optical
disk and a vertical scale height which is smaller than that of the
young, massive stars.  The radius of the HI disk is usually much
larger than that of the optical disk (up to 5 times the Holmberg
radius), and the atomic gas is also less concentrated to the disk in
the vertical direction \citep{K87}.  

If the mass surface density in a region exceeds a certain critical density, 
massive star formation is triggered \citep{K89}.  
So, naturally, the study of interstellar gas, and especially the molecular
gas phase, plays a key role in understanding the structure and
evolution of galaxies.  Unfortunately, direct studies (in emission) of
the cold ($\sim$10 K) H$_2$ gas in molecular clouds are very difficult,
since the energy of the first excited rotational level is of the order
of 500\,K (corresponding to a wavelength of 28\,$\mu$m), and
therefore much larger than the average kinetic temperatures in the
clouds.  Carbon monoxide (CO) is the second most common molecule with
a relative abundance with respect to H$_2$ of about 10$^{-4}$.  The
heavy, asymmetrical CO molecule has rotational levels which are
easily excited by collisions with H$_2$, and it radiates at mm
wavelengths where observations can be carried out using radio telescopes. 
However, the conversion factor which is used to derive the H$_2$
column density from the velocity-integrated CO intensity, $X_{\rm CO}$, is
not well-known, and this is a source of major uncertainty
\citep{YS91,C91}.

CO emission at mm wavelengths has been detected in many galaxies ranging in
distance from the nearby irregular galaxies The Large and The Small Magellanic
Clouds to quasars at a redshift of more than $z$\,$\sim$\,5, and in essentially
all galaxy types.  A large survey of 300 galaxies (of which 236 were
detected) of different types was done by \citet{YXT95} using the FCRAO
14\,m telescope.  The analysis of the data is still ongoing, but some
interesting results have been published, e.g., the conversion factor
($X_{\rm CO}$) appears relatively constant, at least within each galaxy type
\citep{YR01}, and the global star formation efficiency decreases with
increasing galaxy size \citep{Y98}.  In this survey a number of
galaxies were also mapped and radial distributions of the molecular
gas can be examined.  Recently, in a more detailed study, 28 nearby
galaxies were mapped with the 45\,m Nobeyama telescope
\citep{KNN00,NNK01}.  Among other things, this showed that barred and
non-barred galaxies differ in both the distribution and the kinematics
of the molecular gas. Another large survey is the BIMA SONG project 
\citep{RTH01,HTR03}, in which 44 nearby ($v_{\sun} < 2000$ km s$^{-1}$) galaxies 
have been observed with the BIMA array and the NRAO 12m telescope.
Early results show that many galaxies have multiple peaks at the nucleus, 
that the CO distribution usually cannot be fitted with a single exponential
disk model, and that the ratio of the scale length in the CO disk and
the scale length in the stellar disk is on average 1 (but the scatter 
around this average is large).

Detailed morphological studies of selected galaxies are more rare, partly because
there are so few nearby galaxies for which the limited angular
resolution of radio telescopes gives an adequate spatial resolution.
\citet{NKH94} outlined a number of important observational conditions
that should be fulfilled in order to be able to extract useful
information on the properties of the molecular gas in different
regions of a galaxy: (1) a spatial resolution better than one kpc, which is
the typical width of molecular spiral arms and the spacing between
them, (2) high sensitivity in order to detect the weak inter-arm
emission, (3) mapping of a large field in order to get statistically
meaningful results, and (4) filled aperture telescope data, since it
is important to detect the diffuse emission in the inter-arm regions. 
To this list we add a fifth condition, (5) the
observation should be carried out in (at least) two lines per molecular species
(and preferably also partially observed in a few
isotopomers) in order to gain insight into the density, temperature,
and chemistry of the molecular gas.

Single-dish measurements are therefore essential to our determination of
the molecular gas distribution, its relation to star formation
activity, its location with respect to dynamical resonances, and the
total reservoir of gas capable of forming future generations of stars. 
The most detailed single-dish maps of molecular gas in spiral
galaxies are the mappings of M\,51, using the IRAM 30m telescope
\citep{GGC93} and the Nobeyama 45m telescope \citep{NKH94}, and two
projects currently underway to map our neighbor M\,31
\citep{LDH99,NNG00}.  In M\,51 the molecular spiral arms were resolved, and 
streaming motions were detected.  Also, the CO emission
was found to peak on the inside of the spiral arms, giving support to
a scenario where molecular gas agglomerates and forms new stars, which
in their turn produce H$\alpha$ emission and other star formation tracers,
close to the spiral arms.  Support was also found for the proposal that
gravitational instability in the disk is the main formation mechanism
of GMAs (Galactic Molecular Associations of mass $\sim$10$^7$\,M$_{\odot}$).
Other large spiral galaxies that have been completely surveyed in CO with a
single-dish telescope are IC\,342 \citep{CTH01}, M\,101 \citep{KSW91},
NGC\,253 \citep{HWK97}, and NGC\,6946 \citep{TY89}.  Some large scale CO
mapping projects, such as the OVRO maps of M\,51 \citep{AHS99} and
M\,83 \citep{RLH99}, have been done using mm-wave interferometers. 
Although this gives a superior angular resolution and important unique
information, a significant part of the emission is resolved out
(especially the diffuse component in the inter-arm regions).

We have mapped the \COo\ and \COt\  emission in the barred,
grand-design, spiral galaxy M\,83 located at a
distance of 4.5\,Mpc.  The observations were done with the 15\,m
Swedish-ESO Submillimetre Telescope (SEST) providing beam widths
of 45\arcsec\ (\mbox{1--0}) and 23\arcsec\ (\mbox{2--1}), which
corresponds to 980\,pc and 500\,pc, respectively, at the adopted distance. 
M\,83 is among the brightest galaxies in terms of CO emission, and it is
also one of the most nearby barred spirals.  Its low inclination
($\approx$24\dg) and proximity make it a perfect candidate to
investigate the correlation between various tracers of stellar
activity.  

In this paper we present the \COo\ and \COt\ observational data.  We
show the distribution of the molecular gas, as inferred from the CO data,
and compare it to the distributions of optical light, obtained in
different filters, and the HI gas.  We examine the CO
(\mbox{$J=2-1$})/(\mbox{$J=1-0$}) line ratio and its variation over
the disk.  In two forth-coming papers we will present and discuss the
kinematics of the molecular gas, and discuss the relation between the
molecular gas and star formation.

\section{The galaxy M83} 

M\,83 is a relatively nearby, barred, grand-design, spiral galaxy (see
Table \ref{general}).  It is viewed almost face-on and the bar is
aligned with the major axis.  Color images show clumpy, well-defined
spiral arms, evidently rich in young blue stars.  It is also fairly
symmetrical and has no nearby massive optical companions and no
immediate evidence of interaction or outflows.

\begin{table}
\caption{General Parameters of M\,83.}
\label{general}
\begin{tabular}{ll}
\hline
\hline
Morphological Type$^a$	&	SAB(s)c		\\
IR center:$^b$		&			\\
R.A.~(J2000)		&$13^{\rm h}37^{\rm m}00\fs8$		\\
Decl.~(J2000)		&$-29\degr51\arcmin56\arcsec$ 	\\
LSR systemic velocity (opt)$^c$&	506 km s$^{-1}$		\\
Distance$^d$		&	4.5 Mpc	\\
Position angle$^c$	&	45\degr\\
Inclination$^f$		&	24\degr		\\
Holmberg diameter (D$_0$)$^e$&	14\farcm6		\\
$M_{\rm HI}$$^e$		&7.7$\nine$\\
$M_{{\rm H}_2}$$^g$		&3.9$\nine$\\
\hline
\noalign{\smallskip}
\noalign{$^a$ \citet{DDC76};
$^b$ \citet{SW94}; 
$^c$ \citet{C81}; 
$^d$ \citet{TTS03};
$^e$ \citet{HB81}; 
$^f$ \citet{TJD79};
$^g$ this paper}
\end{tabular}
\end{table}

Due to its favorable properties it is one of the most well-observed
galaxies: HI
\citep{RLW74,HB81,TA93}, radio continuum \citep{O85,NBS93}, CO (see
Table~\ref{coobs}), IR continuum \citep{AAW87,FHA92,ECW98}, optical continuum
\citep{JTD81}, Balmer lines \citep{C81,DPD83,TA93}, and in X-ray emission
\citep{IVE99,SW02}.

\begin{table*}
\caption{Previous CO observations of M\,83 divided into
	two groups: single-dish and interferometer observations.
	If no location is given, the map is centered on the nucleus
	and oriented along the major axis.}
\label{coobs}
\begin{tabular}{llll}
\hline
\hline
Reference	&Transition	&Beam and Telescope	&Map description \\
\hline
\citet{CEL78}	&CO(1-0)	&64\arcsec, NRAO 11m	&10 positions, 6\arcmin$\times$3\arcmin	\\
\citet{L87}	&CO(1-0)	&45\arcsec, FCRAO 14m	&21 positions, 5\arcmin$\times$5\arcmin \\
\citet{WRH90}	&CO(1-0)	&45\arcsec, SEST 15m	&196 positions, SE quadrant, 3\arcmin$\times$2\arcmin \\
		&CO(2-1)	&23\arcsec, SEST 15m	&143 positions, SE quadrant, 1\farcm5$\times$1\farcm5 \\
\citet{HNS90}	&CO(1-0)	&16\arcsec, NRO 45m	&85 positions, 3\farcm5$\times$1\arcmin  \\
\citet{W91}
		&CO(2-1)	&22\arcsec, JCMT 15m	&43 positions\\
		&CO(3-2)	&22\arcsec, CSO 10m	&8 positions\\
\citet{PW98}	&CO(3-2)	&14\arcsec, JCMT 15m	&81 positions, 0\farcm7$\times$0\farcm7 \\
			&CO(4-3)	&12\arcsec, JCMT 15m	&46 positions, 0\farcm7$\times$0\farcm5 \\
\citet{IB01} 	&CO(2-1)	&21\arcsec, JCMT 15m	&49 positions, 1\farcm2$\times$2\farcm0  \\
            	&CO(3-2)	&14\arcsec, JCMT 15m	&55 positions, 1\farcm2$\times$1\farcm7 \\
            	&CO(4-3)	&11\arcsec, JCMT 15m	&20 positions, 0\farcm5$\times$0\farcm5 \\
\citet{CTB02}
			&CO(1-0)&55\arcsec, NRAO 12m	&On-the-fly, 10\arcmin$\times$10\arcmin  \\
			&CO(2-1)&28\arcsec, NRAO 12m	&On-the-fly, 8\arcmin$\times$8\arcmin  \\
Thuma et al.~(in prep.)&CO(3-2)	&23\arcsec, SMT 10m	&374 positions, 5\farcm5$\times$2\farcm5 \\
			&CO(4-3)	&17\arcsec, SMT 10m	&21 positions \\
This paper
			&CO(1-0)&45\arcsec, SEST 15m	&1900 positions, 10\arcmin$\times$10\arcmin  \\
			&CO(2-1)&23\arcsec, SEST 15m	&2574 positions, 10\arcmin$\times$10\arcmin  \\

\hline
\citet{LK91}	&CO(1-0)	&12\farcs4$\times$5\farcs4, OVRO	&Inner eastern arm, 1\farcm5$\times$1\farcm5 \\
\citet{KL91}	&CO(1-0)	&8\farcs9$\times$5\farcs8, OVRO	&Western bar end, 1\farcm5$\times$1\farcm5 \\
\citet{HIK94}	&CO(1-0)	&12\arcsec$\times$6\arcsec, NMA	&Nucleus, 1\farcm1$\times$1\farcm1 \\
\citet{RLH99}	&CO(1-0)	&6\farcs5$\times$3\farcs5, OVRO	&Inner eastern arm, 2\farcm9$\times$1\farcm1\\
\hline
\end{tabular}
\end{table*}


The distance to M\,83 is a long-standing issue.
\citet{ST74} used a relation between the luminosity class and the 
angular size of the three largest H{\sc ii} regions to deduce a distance
of 8.9 Mpc, while \citet{D79} used a number
of methods to derive a distance of 3.7 Mpc. 
Recently, Cepheids have been observed in M\,83 by \citet{TTS03} using the VLT.
Twelve Cepheid candidates were observed and the distance was estimated to 
be 4.5$\pm$0.3 Mpc. This distance is also supported by studies that seem to 
indicate that M\,83 interacted with NGC\,5253 some 1--2\,$\times$\,10$^9$ 
years ago \citep{KS95,CCG99}.  This small, metal-poor dwarf galaxy about 
2\degr~SE of M\,83 has roughly the same systemic velocity, and shows a highly
disturbed HI velocity field and intense star formation.  There has
been a number of distance measurements to NGC\,5253
\citep{SSL95,PSS00,GSF00}, and they all favor a distance of about 4 Mpc.

%

\section{Observations and data reduction} \subsection{Observations}
\label{obser}

The CO($J=1-0$ and $J=2-1$) observations of M\,83 were done using the
15\,m SEST\footnote{The Swedish-ESO Submillimetre Telescope is operated
jointly by ESO and the Swedish National Facility for Radio Astronomy,
Chalmers University of Technology} during two epochs,
1989-1994
and 1997-2001 [a description of the
telescope is given by \citet{BDH89}].  Between 1988 and 1994 we observed
\COo\ in the inner disk of M\,83.  A segment of these observations were
presented in \citet{WRH90}.  In 1998 and 1999 we completed this map by
observing the outer disk.  During the latter observations it was
possible to do simultaneous observations in two frequency bands.  We
used this option to observe the \COt\ line in parallel with the
\COo\ line.  The map spacing (11\arcsec ) and integration times were set
by the requirements on the \COo\ data.  In 2000 and 2001 our
primary task was to fill in the central area of the \COt\ map.  For
these observations we chose a grid spacing of 7\arcsec .  We used the
other receiver to observe the $^{13}$CO(\mbox{$J$=1--0}) line, but these
data will be presented elsewhere.

The receivers were centered on the \COo\ and \COt\ lines (115.271 and
230.538 GHz, respectively), where the full half-power beam widths
(HPBWs) are 45$\arcsec$ and 23$\arcsec$, respectively.  Observations
performed during 1988-1994 were made with a cooled Schottky-diode
mixer receiver, while the subsequent observations were done with
SIS receivers.  All receivers operated in the single-sideband mode, and
the average system temperatures, corrected for the atmospheric and
antenna ohmic losses, were 660\,K and 390\,K for the \COo\
observations in the first and second epoch, respectively, and 340\,K
for the \COt\ observations.
The backends were acousto-optical spectrometers, with channel
separations of 0.69 MHz (1.8 and 0.9 km\,s$^{-1}$ at the two rest
frequencies) and total bandwidths varying between 0.5 and 1\,GHz.

We used the dual beam switching mode with a throw of
$\approx$12$\arcmin$ in azimuth.  In dual beam switching the source is
placed first in the signal beam and then in the reference beam.  The
two spectra produced are then subtracted to generate a spectrum with
a minimum of baseline variation.  Great care was taken to ensure the
quality of the data.  No observations were done below the elevation of
30\degr~or above 84\degr~(M\,83 passes within 1\degr~from zenith at
SEST), and pointing was regularly checked and updated using the
SiO($v$=1, \mbox{$J=2-1$}) maser line of the nearby AGB-star W\,Hya
($\approx$3\dg~from M\,83).  Total pointing errors were typically less
than 3\arcsec , and data with pointing offsets suspected to be larger
than 4\arcsec~were disregarded in the subsequent data reduction (when
possible, map points with such data were reobserved).  The intensity
scale was calibrated using the conventional chopper wheel method, and
the internal calibration errors in the corrected antenna temperatures, 
$T_{\rm A}^*$, is within $\pm$10\% according to the SEST manual.
In this paper we use the main beam
brightness temperature scale $T_{\rm mb}$, which is defined by $T_{\rm
mb} \equiv T_{\rm A}^*/\eta_{\rm mb}$.  According to the SEST manual
the main beam efficiencies [$\eta_{\rm mb}$(115 GHz)=0.7, $\eta_{\rm
mb}$(230 GHz)=0.5] have been constant throughout the period of our
observations.

\begin{table*}
\caption{SEST observation parameters for M\,83.}
\label{obstable}
\begin{tabular}{lll}
\hline
\hline
				&CO(J=1--0)	&CO(J=2--1)\\
\hline
Center of the CO map	&\multicolumn{2}{l}{R.A. = $13^{\rm h}36^{\rm m}59\fs 4$  (J2000)} \\
			&\multicolumn{2}{l}{Decl.= $-29\degr 52\arcmin05\arcsec$ (J2000)} \\
Rest frequency          & 115.271204 GHz         & 230.53799 GHz \\
Beam size 			&45\arcsec &23\arcsec\\ 		
Center velocity (LSR)           &520 km s$^{-1}$&520 km s$^{-1}$ \\
Average rms ($T_{\rm mb}$ scale)&74 mK		&90 mK 		\\
Channel separation		&$\leq$0.69 MHz &0.69 MHz\\
Number of channels		&$\geq$700 	&1000\\
Main beam efficiency ($\eta_{\rm mb}$)	&0.7 	&0.5\\
Aperture efficiency ($\eta_{\rm A}$)	&0.58 (27 Jy/K) &0.38 (41 Jy/K)\\
\hline
\end{tabular}
\end{table*}

The observations were centered on the coordinates $13^{\rm h}36^{\rm
m}59\fs 4$, $-29\degr 52\arcmin05\arcsec $ (J2000), which is the
optical center given by \citet{DDC76}.  For \COo\ we have 3634
spectra in 1900 positions with 11\arcsec~spacing, and the coverage is
complete out to a radius of 4\arcmin 20\arcsec .  The grid extends
parallel to the equatorial coordinate system, and the dense spacing
(1/4 of the HPBW) was chosen in order to facilitate the use of
deconvolution techniques.
The integration time per position was chosen to render an rms around
70\,mK ($T_{\rm mb}$-scale) at the original velocity resolution (1.8 km
s$^{-1}$).  During normal weather conditions we had a typical
on-source integration time of 60 seconds. Among the final 1900
spectra we have $\ge$\,3$\sigma$ detections in 1201 positions, 
i.e. a detection rate of 63\%.
The average rms noise in our data set is 74\,mK, which should be
compared to the peak temperature which usually lies around 0.3\,K
($T_{\rm mb}$-scale) in the spiral arms in the outskirts of the map.

In \COt\ we have two maps with different spacings covering two,
partly overlapping, regions.  The inner 5\arcmin$\times$3\arcmin~is covered
with a 7\arcsec\ grid, while the rest of the optical disk (out to a
radius of 4$\farcm$5) is covered with an 11\arcsec\ grid. In total,
we have spectra in 2574 positions, out of which we have 
$\ge$\,3$\sigma$ detections in 1898 positions (73\% detection rate).
The spectra have an average rms of about
90\,mK at the original velocity resolution (0.9\,km\,s$^{-1}$).
The observational parameters are summarized in Table~\ref{obstable}.

\subsection{Data reduction}

The data reduction was carried out with {\sc class}\footnote{Continuum and
Line Analysis Single-dish Software from the Observatoire de Grenoble
and Institut de Radio-Astronomie Millim\'etrique }, {\sc
drp}\footnote{Data Reduction Package developed at Onsala Space
Observatory}, and {\sc xs}\footnote{{\sc xs}: Spectral Line Reduction
Package for Astronomy developed at Onsala Space Observatory}.  During the
\COo\ observations, the number of channels and the frequency resolution
varied between the spectrometers.  
We therefore resampled all data to 1.8\,km\,s$^{-1}$ (0.69 MHz) and
by selecting the central 700 channels this gave us a (fully sufficient) 
total bandwidth of 1260\,km\,s$^{-1}$.

After flagging bad channels, and replacing them with a value obtained from
interpolation between adjacent channels, all spectra at the same
position were averaged using a weight determined by the noise level.
The baselines were stable, and only linear polynomial fits were
subtracted.  From the spectra we created a {\sc fits} data cube for
each transition, where the first two axes represent spatial coordinates, and
the third axis represents frequency (or velocity).

We adopted a relatively dense grid spacing, compared to the beamwidth, in
order to be able to increase the spatial resolution of the data using
a {\sc mem}-deconvolution routine (maximum entropy method); the
{\em Statistical Image Analysis} (SIA) routine \citep{R00}.
This procedure is particularly effective on a data
set as densely spaced as ours (spacing in the range 1/4--1/2 of the
HPBW), due to the high redundancy.  The routine starts by
deconvolving a velocity-integrated intensity map, where the velocity
range is chosen to be large enough to encompass all the emission.
Once this is done, the procedure splits the initial velocity range 
into three equally wide regions, calculates their velocity-integrated
intensities, and then deconvolves these three maps, using the
deconvolved map from the previous step as the input guess.  The
deconvolution continues in this hierarchical manner, successively
narrowing the velocity interval, until a velocity range of
5\,km\,s$^{-1}$ is reached.  To deconvolve beyond this point is not
meaningful since the S/N-ratios in the (narrow velocity range) maps
drop below a critical value.  
For each of the two CO transitions we created 81 (3$^4$) maps per transition
(each covering about 5\,km\,s$^{-1}$) which we assembled into a data
cube per transition.  By convolving the {\sc mem}-deconvolved data cube by a
Gaussian with a FWHM equal to the observational HPBW and comparing
with the raw spectra, we verified that the {\sc mem}-results are
reliable, and that the raw data set is homogeneous, i.e., free from
variations in gain and pointing errors.  The angular resolutions of
the {\sc mem}-cubes are estimated to be $\approx$22\arcsec\ and
$\approx$14\arcsec\ for the \COo\ and \COt\ data, respectively, and the
velocity resolution is 5\,km\,s$^{-1}$.

We have also constructed convolved data sets by convolving the raw
data cubes with a HPBW beam size of 20$\arcsec$ and 15$\arcsec$ for the \COo\
and \COt\ data, respectively (the spacing in these cubes are 11\arcsec ).
Even if this degrades the resolution by $\approx$10--20\%, the average rms 
noise in these cubes decreases from 74 mK to 25 mK and from 90 mK to 31 mK, 
respectively. The dramatically decreased noise level is
due to the fact that the spacing is small with respect to the 
HPBW of the convolution kernels.
Since the peak intensity in the spectra are relatively unaffected by this process, 
the average S/N-ratio in the spectra increases by a factor of $\approx$3.
As a result of this, the detection rate increases to 92\% and 89\% for the 
\COo\ and \COt\ data, respectively (cf Sect. \ref{obser}).
These cubes were used to cross-check the
results we got from the {\sc mem}-deconvolved data.  Some results were also
derived directly from the convolved cubes, especially in cases where
the velocity resolution was more important than the spatial
resolution.  Furthermore, we convolved the raw \COt\ cube with a beam
size of 44\arcsec\ and regridded it to 11\arcsec\ spacing in order to
compare with the convolved \COo\ cube at the same resolution
($\approx$49\arcsec ).

In order to avoid problems with baseline irregularities when calculating
the zero moment (over velocity) of the intensities we used 
a sliding window technique where at each position integration was
performed inside a velocity window whose center and width depend on
the position within the galaxy.  We checked carefully that the window
always covered all the emission.  

\subsection{Correction for the error beam}
\label{errorbeam}
A closer inspection of the \COt\ spectra reveals, apart from the expected
main component seen in the \COo\ spectra, a broad (velocity width
400\,km\,s$^{-1}$) and faint (peak intensity around 0.03\,K) component.
Despite the low peak intensity of this feature, its integrated intensity
is quite appreciable due to its large width.  Typically, it makes up
to 20\%--35\% of the total integrated intensity.  This feature was
discovered in the {\sc mem}-deconvolved data, which are noise-free,
but it can be seen also in spectra produced by convolving the raw
\COt\ data cube with a 45\arcsec\ beam.  The appearance of this broad
component changed with position in the disk, and we found that its
shape could, in any given position, be reproduced very well by
convolving the data set with a relatively large Gaussian beam.  This
indicates that the SEST picks up emission from a much larger area than
the HPBW of 23", i.e., at 230\,GHz the SEST suffers from a
non-negligible error beam.

We have corrected the \COt\ data sets
for this error using a method described in Appendix \ref{apperrbeam}.
We have found no trace of an error beam contribution in the \COo\ data
(not unexpectedly, considering the higher main beam efficiency), and
consequently we have not applied any correction to these data.


\section{The data} \subsection{Spectra}

In general, the spectra have a smooth and symmetric shape, but in some areas
they show deviations.  The spectra close to the nucleus are asymmetric,
show extended wings, and a few of them have multiple peaks.  
The main reason for this is a large velocity gradient
over spatial scales small compared to the telescope beam.
Further out in the disk there are a few (interarm) regions which show
multiple-peak spectra, such as the ``Gould-belt structure'' located
about 3$\arcmin$ SE of the galaxy center \citep{C01}.  However, in the
vast majority of the observed positions the spectral feature can be
well described by a single, Gaussian profile.

\subsubsection{The \COo\ spectra}

In the arms we find that the peak temperatures are typically 0.2--0.4\,K, and
in the inter-arm regions they are about 0.1\,K. The spectra are widest in the
nuclear region with (Gaussian) widths of 80--100\,km\,s$^{-1}$. 
Further out the linewidths are of the order of 7--14\,km\,s$^{-1}$ in
the arms, with no systematic difference between arm and inter-arm
regions.

\begin{figure*}
	\resizebox{0.5\hsize}{!}{\rotatebox{-90}
	{\includegraphics{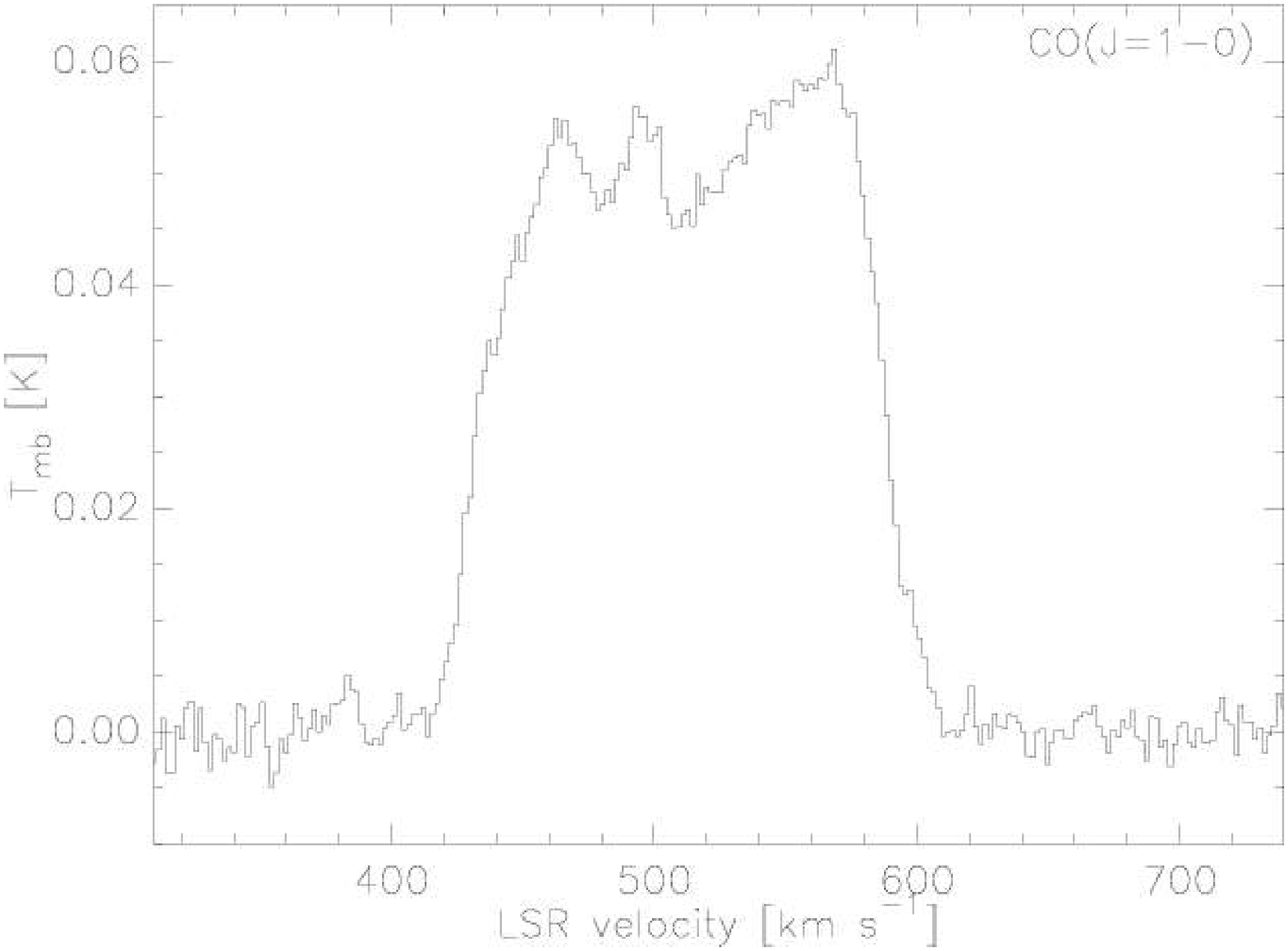}}}
	\resizebox{0.5\hsize}{!}{\rotatebox{-90}
	{\includegraphics{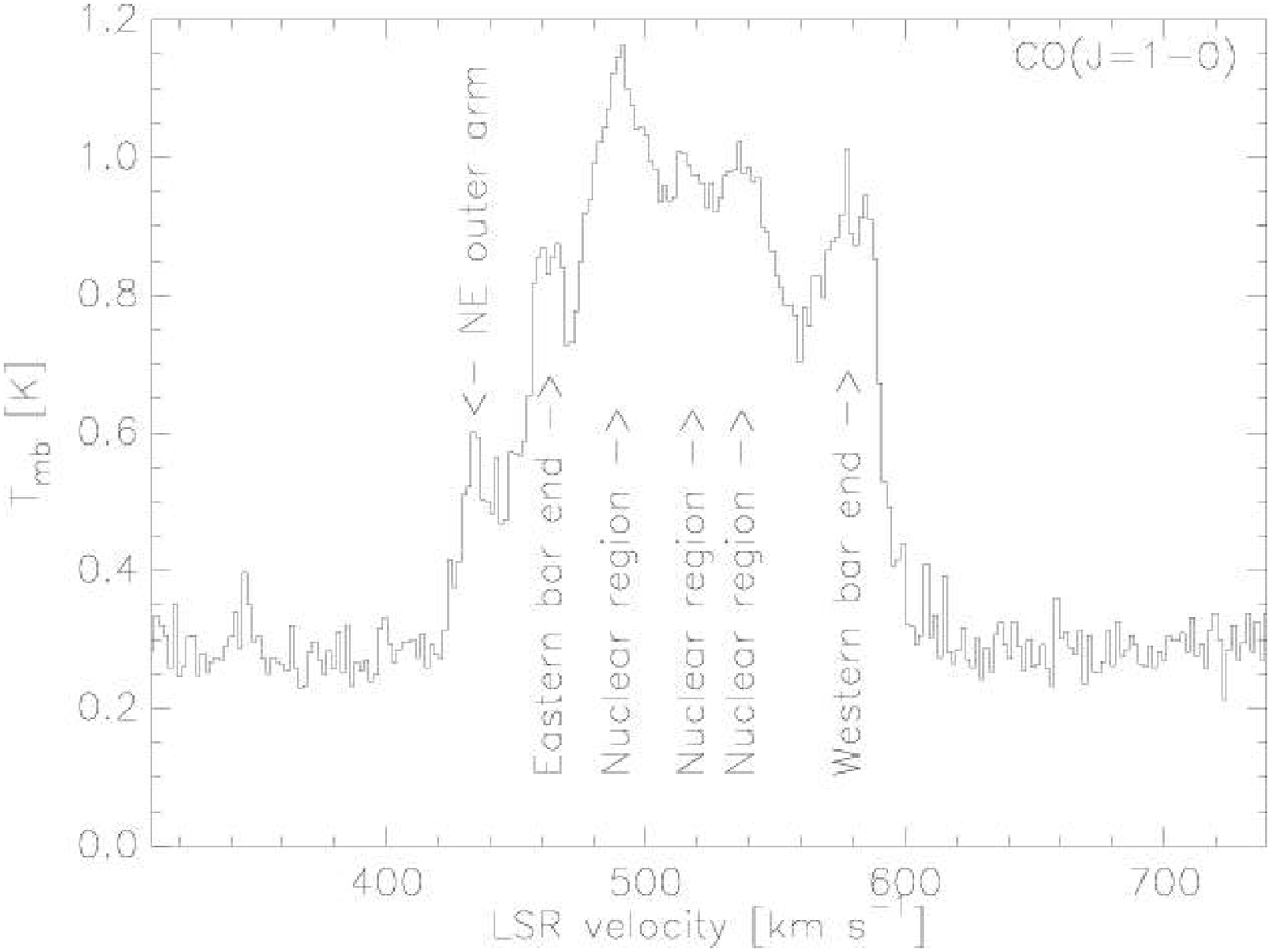}}}
	\caption{Left: Global \COo\ spectrum. Right: Peak
	intensity \COo\ spectrum. }
	\label{intpeak}
\end{figure*}

The global \COo\ spectrum (left panel in Fig.~\ref{intpeak}) has a
velocity-integrated intensity of 8.18$\pm$0.04\,K\,km\,s$^{-1}$
(excluding the calibration error) and a noise level of 1.8\,mK 
(all errors in this paper are 3$\sigma$ unless otherwise noted). The receding
part of the galaxy (SW) has more \COo\ emission than the approaching
part (NE): the velocity interval 510--610\,km\,s$^{-1}$ is 18\% more
luminous than the 410--510\,km\,s$^{-1}$ interval.  An excess of
higher-velocity emission has also been seen in the atomic gas. 
\citet{HB81} found that the higher-velocity part of the H{\sc I}
emission is 16\% more luminous than the lower-velocity part.

The peak intensity spectrum (Fig.~\ref{intpeak}, right panel)
has been obtained by selecting, for
every channel, the highest intensity in the map.  Therefore, the noise
level in this spectrum reflects the noise in the overall worst case. 
From studies of the channel maps we conclude that the peaks in this
spectrum arise from the nuclear region (487, 514, and
534\,km\,s$^{-1}$), the western bar end (580\,km\,s$^{-1}$), the
eastern bar end (460\,km\,s$^{-1}$), and the NE outer spiral arm
(437\,km\,s$^{-1}$).

\begin{figure*}
	\resizebox{\hsize}{!}{\rotatebox{-90}
	{\includegraphics{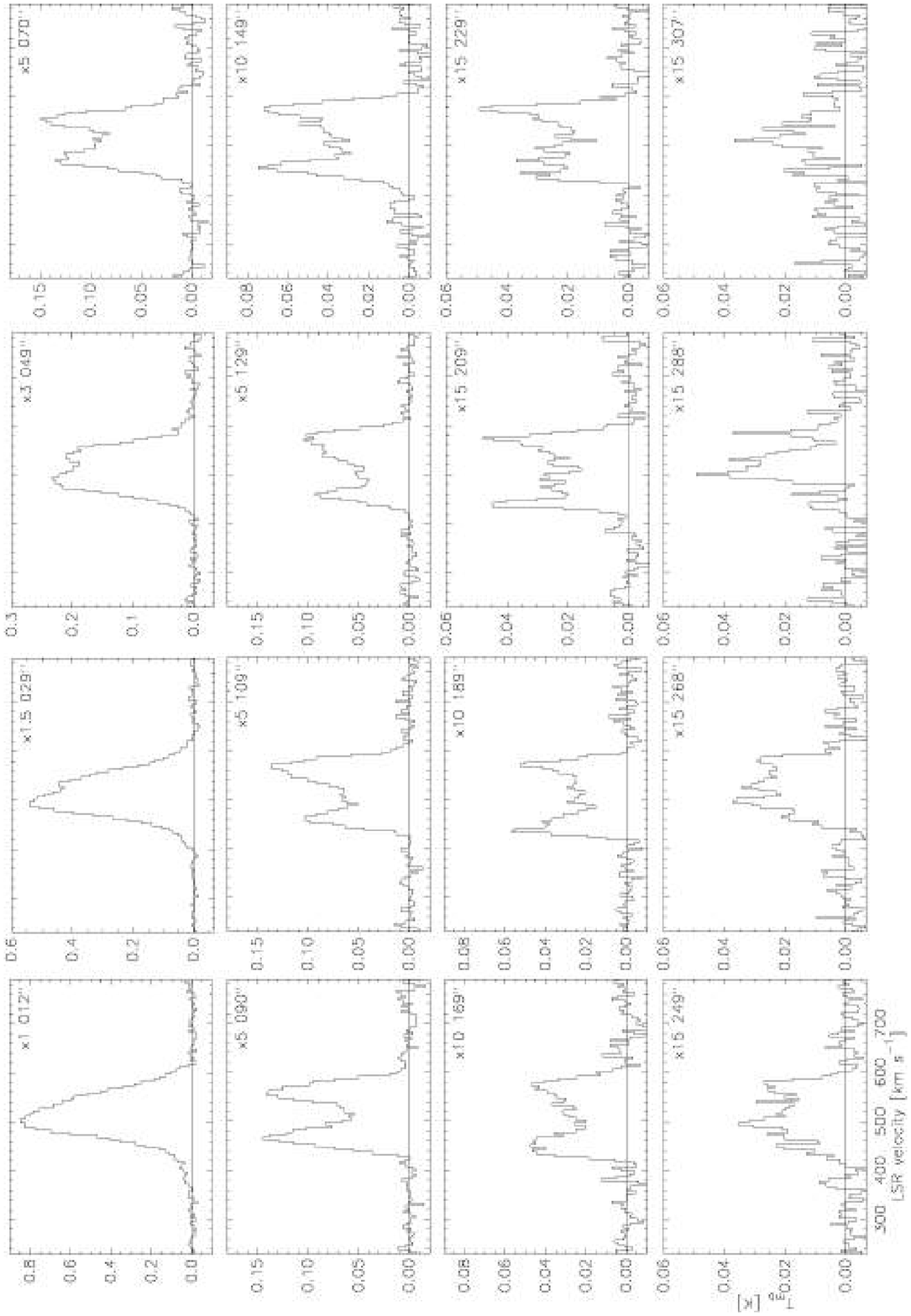}}}
	\caption{\COo\ spectra produced by averaging all spectra
	within successive, concentric (and inclination-corrected) annuli.
	The average radius (in arc-seconds) of each annuli is shown in
	the upper right corner.  The center spectrum is shown in the upper
	left corner, and the outermost spectrum in the lower right corner.
	Note that the intensity scale varies from panel to panel (the scaling
	factor is shown in the upper right corner of each panel). The velocity
	resolution has in this image been reduced to 5 km s$^{-1}$.}
	\label{concentric}
\end{figure*}

By adding the \COo\ spectra (from the raw data cube) in concentric,
20\arcsec~wide annuli (compensated for inclination and position angle
and centered on the IR nucleus, see Table~\ref{general}), we produced
the spectra in Fig.~\ref{concentric}.  They show how the average \COo\
velocity-integrated intensity changes with galactocentric radius (which
is indicated, in arc-seconds, in the upper right corner of each panel). 
The coverage in the map is complete to a radius of 229\arcsec . 

\subsubsection{The \COt\ spectra}

\begin{figure*}
	\resizebox{0.48\hsize}{!}{\rotatebox{-90}
	{\includegraphics{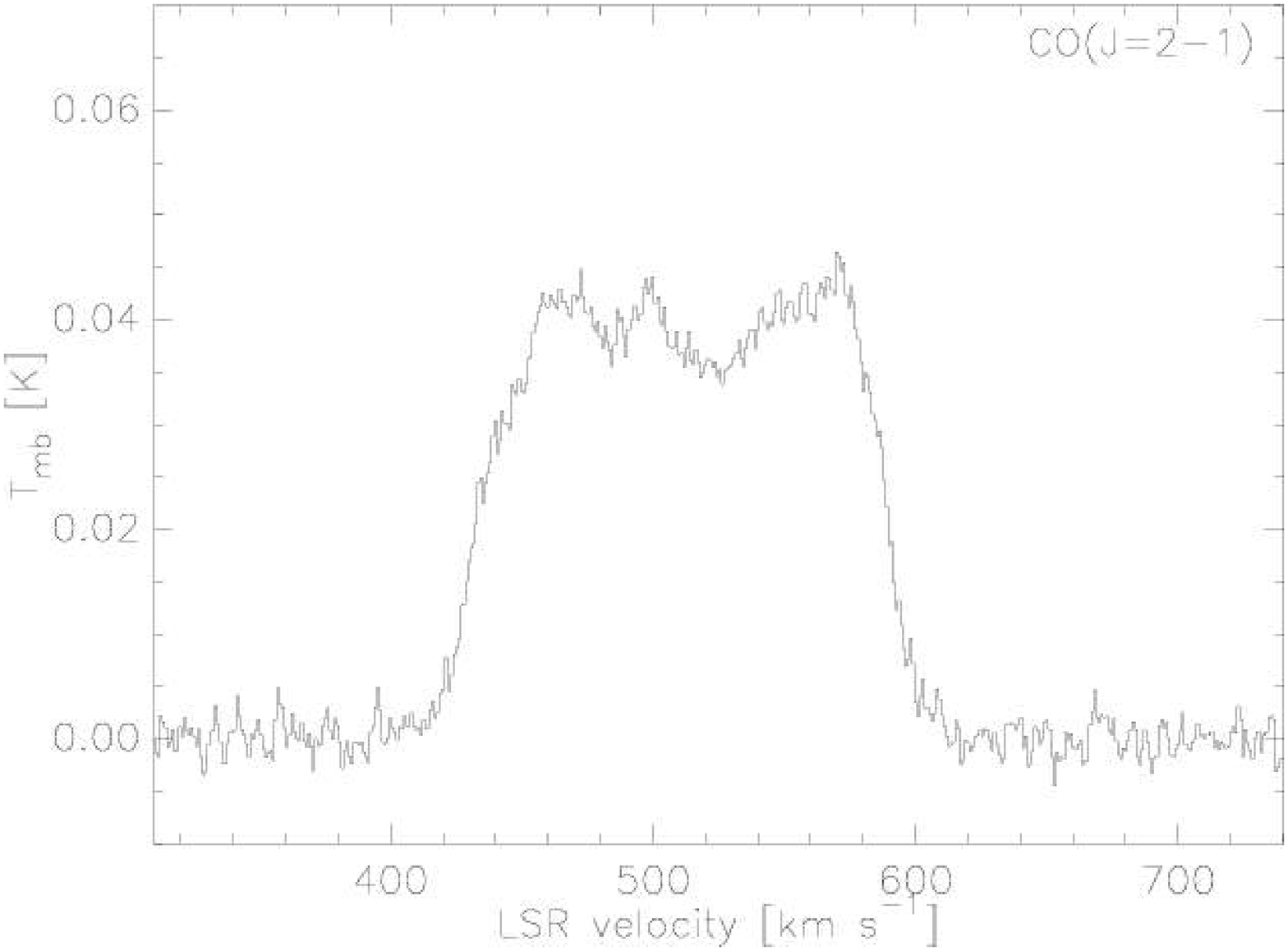}}}
	\resizebox{0.48\hsize}{!}{\rotatebox{-90}
	{\includegraphics{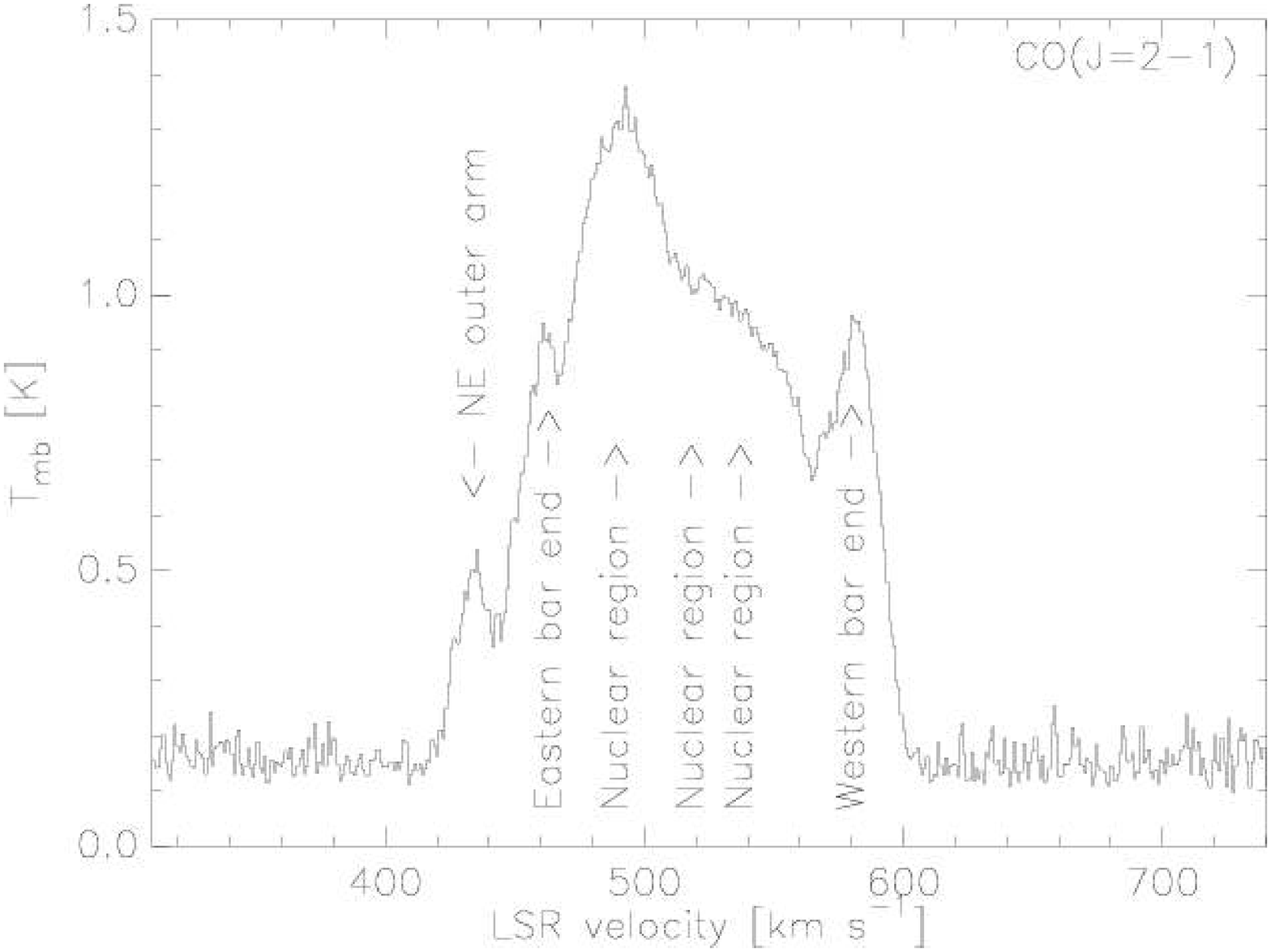}}}
	\caption{{\bf Left:} Global \COt\ spectrum.  {\bf Right:} Peak intensity
	\COt\ spectrum.  Before producing these spectra the \COt\ data cube was
	regridded to 11\arcsec\ spacing. }
	\label{intpeak2}
\end{figure*}

The global and peak intensity \COt\ spectra show characteristics very
similar to those of the \COo\ data, Fig.~\ref{intpeak2}
[since the \COt\ spectra were obtained with two
different spacings, we regridded these data to a common
11\arcsec\ spacing before producing the spectra in Fig.~\ref{intpeak2}].
The main difference is that the peak intensity spectrum is brighter than the
corresponding \COo\ spectrum. This is most 
likely an effect of the emitting objects being smaller than the 
\COo\ beam.

In Table~\ref{comp12} we summarize the characteristics of the global spectra. 
The global line intensity ratio $R_{21}\equiv\int{T_{\rm
mb}(2-1)dv}/\int{T_{\rm mb}(1-0)dv}$ is 0.77$\pm$0.01, based on the
errors from Table \ref{comp12}.  Note that, including errors in the
calibration and the main beam and error beam efficiencies, the result is
0.77$\pm$0.16.
However, it should be noted that the relative calibration of the \COo\ and 
\COt\ data may be better than the absolute calibration and hence the 
error in their ratio can be smaller than indicated by the absolute 
error.

\begin{table}
\caption{Peak intensities, velocity-integrated intensities (with
errors), intensity-weighted velocities, and equivalent widths of
the global \COo\ and \COt\ spectra in Figs.~\ref{intpeak} and
\ref{intpeak2}.  The errors are estimated from the noise levels.
In the column for integrated intensities we also show the absolute errors.
}
\label{comp12}
\begin{tabular}{lcccc}
\hline
\hline
		&Peak	&Integral 	&Centroid 	&Eq. width\\
		&[K]	&[K km s$^{-1}$]&[km s$^{-1}$]	&[km s$^{-1}$]\\		
\hline
\COo\	        &0.061	&8.18$\pm$0.04$\pm$0.82	&515$\pm$1	&134$\pm$1\\
\COt\	        &0.046	&6.28$\pm$0.03$\pm$0.62	&513$\pm$1	&135$\pm$1\\
\hline
\end{tabular}
\end{table}

\subsection{The maps}

\subsubsection{\COo }

\begin{figure*}
	\resizebox{\hsize}{!}{\rotatebox{-90}
	{\includegraphics{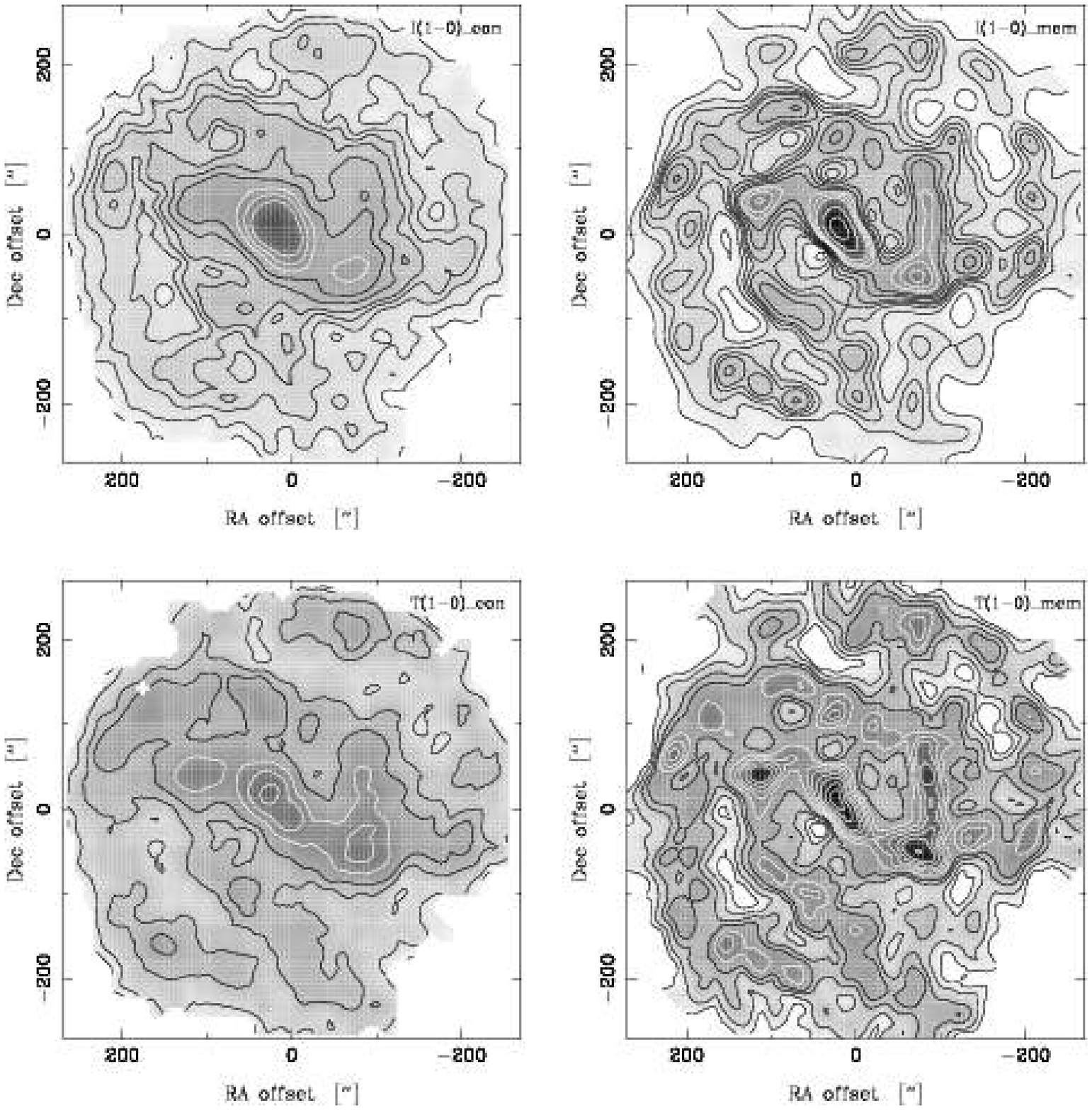}}}
	\caption{The \COo\ maps.  Velocity-integrated intensity in the
	convolved data set (resolution $\approx$49\arcsec ; upper left) and in
	the {\sc mem}-deconvolved data set (resolution $\approx$22\arcsec ; upper
	right).  Contours are: 1, 3, 5, 7, 9, 13, 18, 27 (first white
	contour), 37, 50, 100 and 175\,K\,km\,s$^{-1}$.  Maximum intensities
	in the convolved and {\sc mem}-deconvolved maps are 127 and
	235\,K\,km\,s$^{-1}$, respectively.  Peak intensities in the convolved
	data set (lower left) and in the {\sc mem}-deconvolved data set (lower
	right).  Contours are: 0.05, 0.1, 0.2, 0.3, 0.5 (first white contour),
	0.7, 0.9, 1.2, 1.5, 2.0 and 2.5\,K. Peak intensity at the center are
	1.4 and 2.8\,K for the convolved and deconvolved set, respectively. }
	\label{peakint1}
\end{figure*}

Figure~\ref{peakint1} shows maps of the \COo\ velocity-integrated intensity
(hereafter $I_{1-0}$) and the peak intensity (hereafter
$T_{1-0}$) for both the convolved data set (left column) and the {\sc
mem}-deconvolved data set (right column).
The \COo\ emission is concentrated to the nucleus, the central part of
the bar, the bar ends, and the spiral arms.  The velocity-integrated
intensity at the nucleus and the eastern and western bar ends are: 73,
24, and 30\,K\,km\,s$^{-1}$, respectively, which is very close to the
corresponding values in \citet{CTB02}: 73, 22, and 28\,K\,km\,s$^{-1}$. 
In the {\sc mem}-deconvolved data set these features are brighter due
to the increased spatial resolution: at the eastern bar end we measure
49\,K\,km\,s$^{-1}$, at the western bar end
52\,K\,km\,s$^{-1}$, and at the nucleus the maximum
$I_{1-0}$ is 235\,K\,km\,s$^{-1}$ ($T_{1-0}$=2.8\,K).
By fitting an ellipse to the {\sc mem}-deconvolved $I_{1-0}$-distribution
we find that the CO peak coincides with the IR nucleus within the
absolute position accuracy of our maps (3\arcsec ).
The axes
ratio of this ellipse is 60\arcsec:29\arcsec\ with a position angle
of 36\degr.  In the {\sc mem}-deconvolved maps the spiral arms are
resolved, and they can be followed for almost 360\degr .  The emission
in the spiral arms breaks up into regularly spaced maxima, which we
interpret to be giant molecular associations (GMAs), a phenomenon
previously seen in e.g. M\,51 \citep{NKH94}.  The outer arm, to the SE
of the nucleus, seems somewhat disturbed in the sense that in the {\sc
mem}-deconvolved $T_{1-0}$ map it is possible to discern two parallel
arms. Further toward the west, this arm appear to almost
merge with the inner arm, which also appears to be the case on the 
opposite side of the galaxy (NE of the nucleus).

\begin{figure*}
	\resizebox{0.5\hsize}{!}{\rotatebox{0}
	{\includegraphics{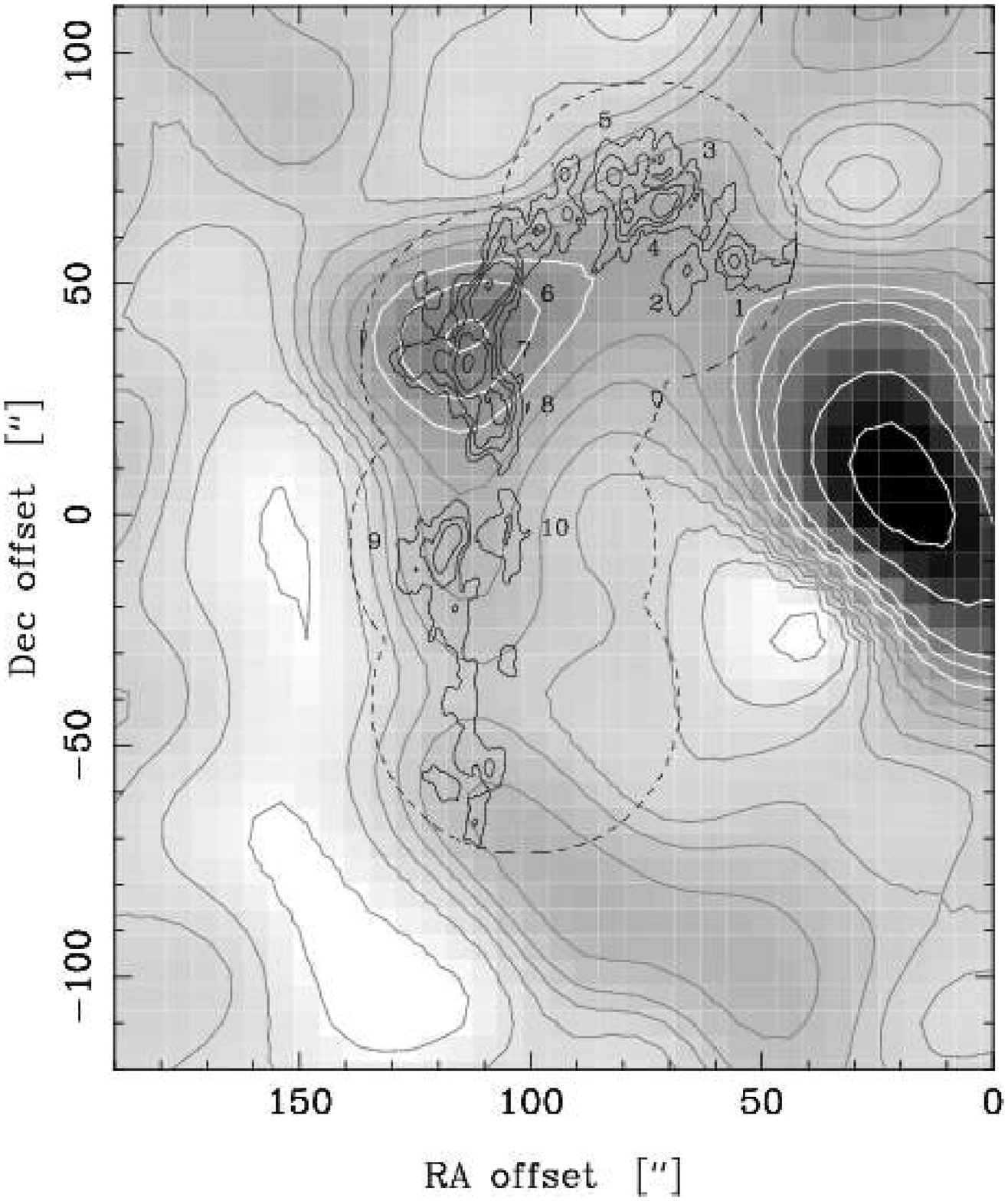}}}
	\resizebox{0.5\hsize}{!}{\rotatebox{0}
	{\includegraphics{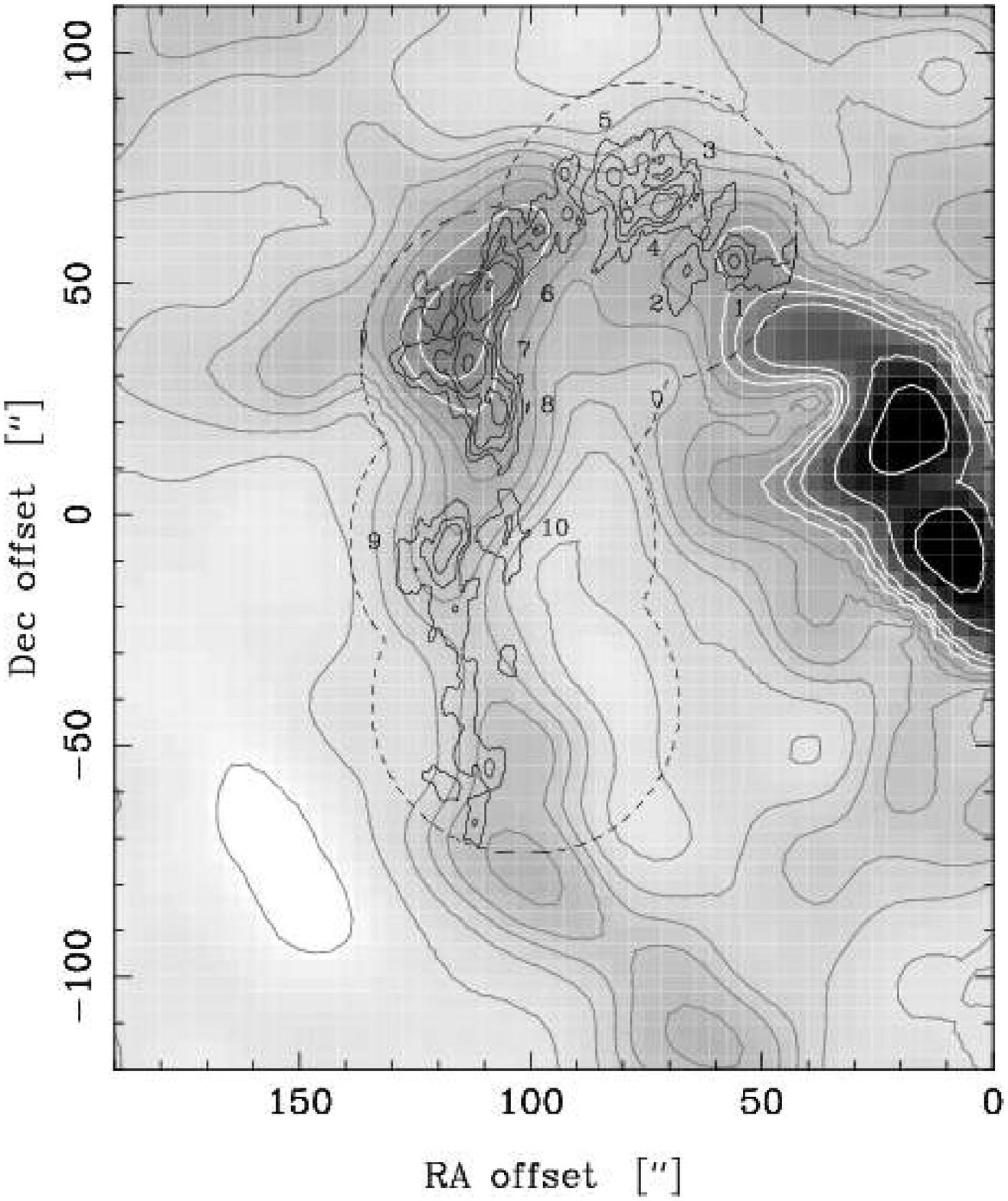}}}
	\caption{OVRO interferometer \COo\ emission (thin lines)
	superposed on a gray+contour map of our \COo\ [left] and \COt\ data [right].
	}
	\label{RLH}
\end{figure*}

\citet{RLH99} (hereafter RLH) mapped the \COo\
emission in the inner eastern arm using the OVRO Millimeter Array
(spatial resolution 6\farcs5 $\times$ 3\farcs5).  In Fig.~\ref{RLH} we
show their data overlayed on our deconvolved $I_{1-0}$ map.  
The components close to the nucleus (components 1--5
using their nomenclature), are not prominent in our data.  They are
distributed along the leading edge of the bar.  The reason is probably
that a shock in this region creates substructures (in temperature, in
density, and/or in velocity) in the CO distribution which elsewhere is
smooth and therefore undetectable by an interferometer (RLH found that 
only 2--5\% of the single-dish emission was recovered in their
interferometer observations). Our eastern
bar end is located between components 6 and 7, and the spiral arm
segment which continues southward is located between components 9 and
10.


\subsubsection{\COt }

\begin{figure*}
	\resizebox{\hsize}{!}{\rotatebox{-90}
	{\includegraphics{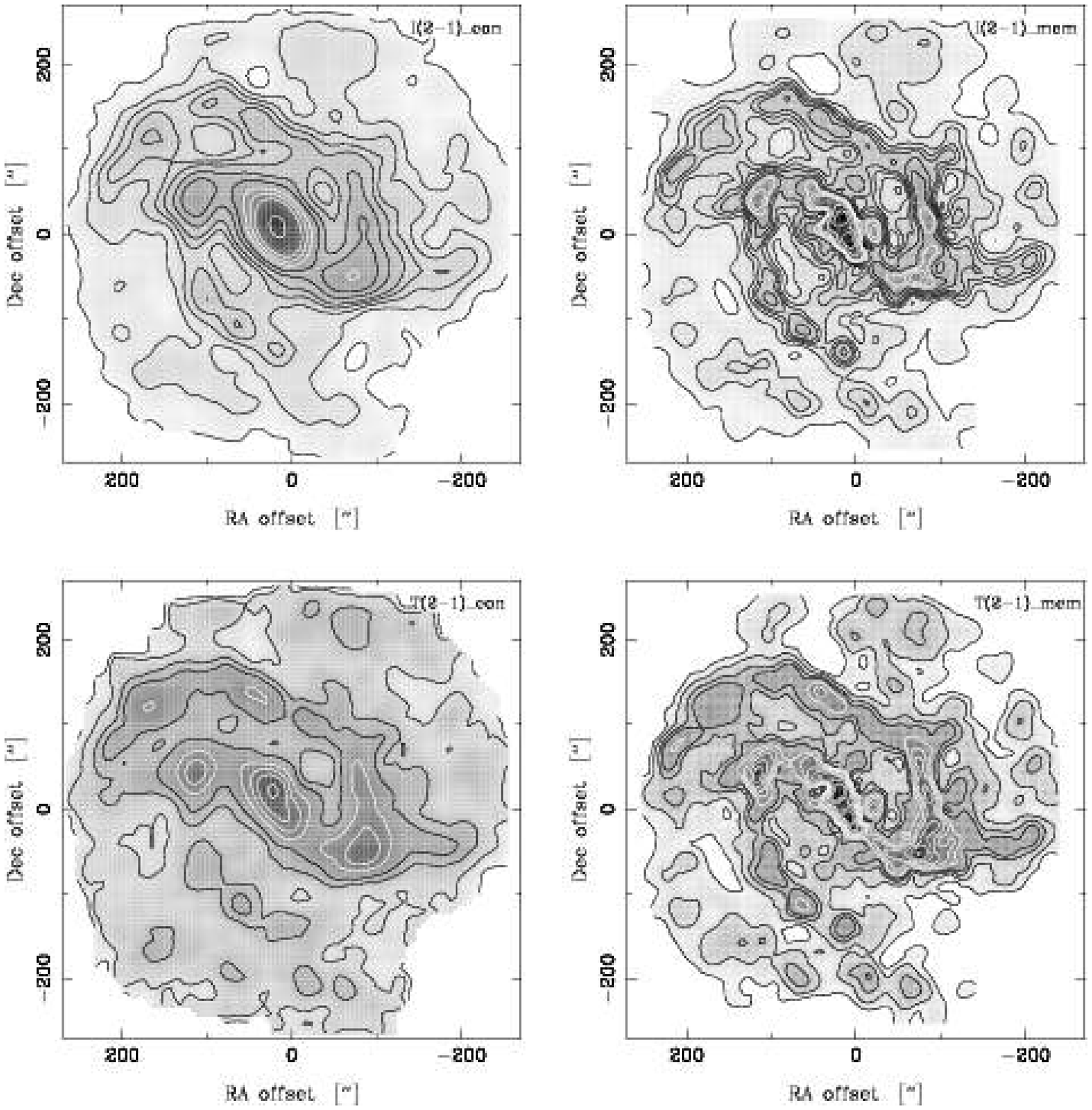}}}
	\caption{
	Same as in Fig.~\ref{peakint1} but for \COt\ . The contours are also the
	same.  The resolution of the maps are 27\arcsec~and 12\arcsec~for the
	convolved and the {\sc mem}-deconvolved data set, respectively.  For
	the convolved data maps the values at the center are
	110\,K\,km\,s$^{-1}$ and 1.3\,K. In the {\sc mem}-deconvolved map, the
	nucleus breaks up into two discrete sources, with similar
	velocity-integrated and peak intensities: 240\,K\,km\,s$^{-1}$ and 3.0\,K.
	}
	\label{peakint2}
\end{figure*}

The \COt\ maps show structures very similar to those in the \COo\ maps,
however with a higher angular resolution, Fig.~\ref{peakint2}.  The
deconvolved $I_{2-1}$ map shows the ``double nucleus'' previously seen in
the CO(\mbox{$J=3-2$}) and CO(\mbox{$J=4-3$}) lines by \citet{PW98}. 
This ``double nucleus'' is apparent also in our raw \COt\ data, but it is
not possible to see in the convolved $I_{2-1}$ map, since this map is
made using data which was both convolved (with a 15\arcsec~beam) and
regridded (to 11\arcsec~spacing).  Since the gray-scale and contour
levels in Figs.~\ref{peakint1} and \ref{peakint2} are the same, we can
directly infer that the \COt\ emission is less concentrated to the
arms than the \COo\ emission.


\section{Results and discussion}
\subsection{Molecular mass in M\,83}
\label{molmass}

The H$_2$ column density in external galaxies is commonly derived from the
velocity-integrated \COo\ intensity ($\ico=\int \Tmb dv$) using the
expression
\begin{equation}
\label{xco}
	N({\rm H}_2)=X_{\rm CO}\times \ico \ ,
\end{equation}
where \XCO~is a conversion factor.  The actual value of this factor,
and its applicability in different environments have been widely
discussed, see e.g.~\citet{W95,AST96,C00a,YR01}. In this paper we have adopted the value \XCO = 2.3$\times
10^{20}$\,(K\,km\,s$^{-1}$)$^{-1}$\,cm$^{-2}$ \citep{SBD88}, which is
supported by the analysis done in RLH. In order to convert the
measured intensities into mass surface density we used the expression
\begin{equation}
\Sigma(\hmm) = 
3.7 \times \ico \times \cos\,i = 3.3 \times \ico {\rm~M_\odot~pc^{-2},}
\end{equation}
where {\em i} is the inclination of the galaxy, and $\ico$ is given in
K\,km\,s$^{-1}$ (This expression can be obtained by integrating Eq. 
\ref{xco}). Using this conversion factor to convert the
integrated intensity of the global \COo\ spectrum
(8.18\,K\,km\,s$^{-1}$, Tab.~\ref{comp12}) we find that the average
(over the entire optical disk, and including the bar and the nucleus)
surface mass density is 27.2\,M$_\odot$\,pc$^{-2}$, and that the total
molecular gas mass within the surveyed area is 3.3$\times
10^9$\,M$_\odot$.

\subsection{The radial distribution}

\begin{figure*}
	\resizebox{0.49\hsize}{!}{\rotatebox{-90}
	{\includegraphics{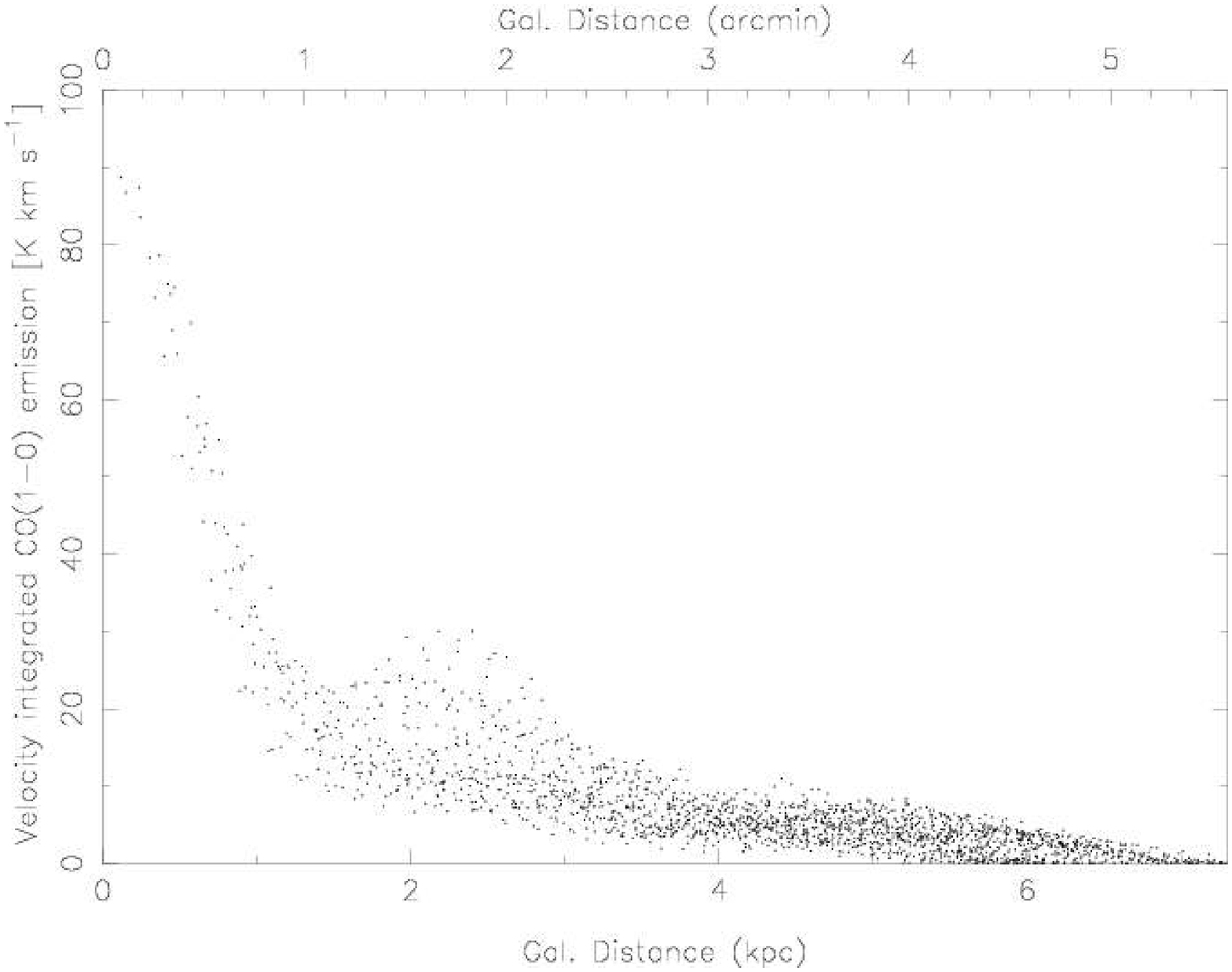}}}
	\resizebox{0.49\hsize}{!}{\rotatebox{-90}
	{\includegraphics{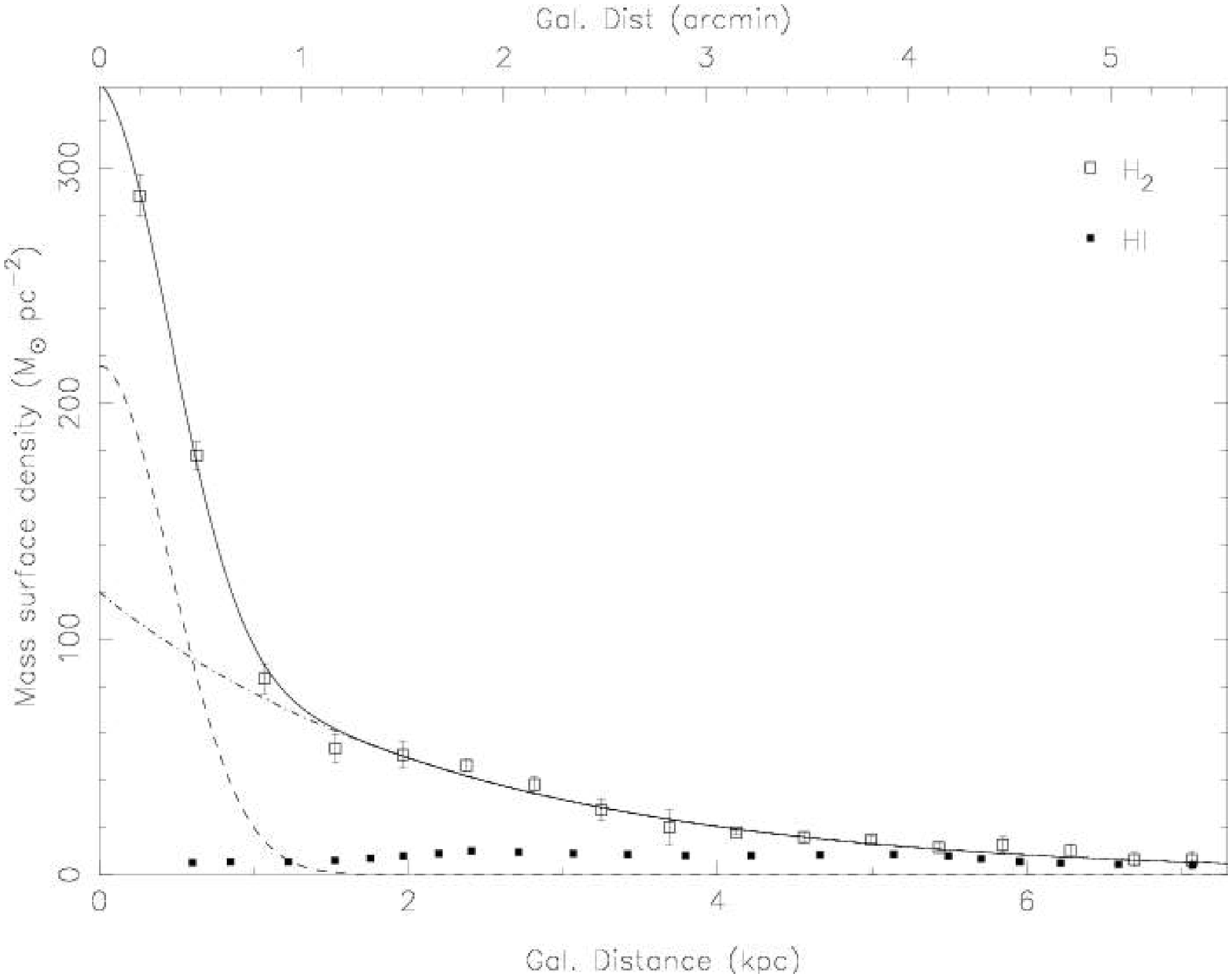}}}
	\caption{The \COo\ velocity-integrated intensity versus galactocentric
	distance is shown in the left panel (obtained from the convolved 
	data set). The dispersion is dominated by the
	arm-interarm contrast, and the bar-ends are evident at $\approx$2 kpc. 
	In the right panel the H$_2$
	mass surface densities are shown with formal
	errors (the data are taken from the spectra in Fig.~\ref{concentric}),
	together with HI mass surface densities. 
	Also shown, as a solid line, is the fit of an H$_2$ mass distribution
	described in the text.  It consists of a Gaussian central part (dashed
	line) and an exponential disk (dash-dotted line).  }
	\label{expdec}
\end{figure*}

\subsubsection{Molecular gas}

We have, using the sliding-window technique described above, calculated
the velocity-integrated intensity, centroid velocity, peak intensity, and 
center and width of a Gaussian emission profile fitted to the spectrum
for all spectra in all data cubes.  The left panel of Fig.~\ref{expdec} shows
the \COo\ velocity-integrated intensities (obtained from the convolved
data set) as a function of the (inclination-corrected) distance from the
galaxy center.  The scatter is dominated by the arm-interarm contrast. 
The data in the right panel of Fig.~\ref{expdec} were obtained from
the integrated intensities of spectra averaged within concentric
annuli (see Fig.~\ref{concentric} and Table~\ref{exptab}) by
converting the \COo\ integrated intensity into mass surface density as
described above.  In this procedure we used the raw data set, since we
needed neither the higher spatial resolution of the {\sc
mem}-deconvolved spectra nor the lower noise of the convolved spectra. 
The error bars reflect the uncertainty in the azimuthally-averaged
radial distribution. To these data we fitted an H$_2$ mass surface density
distribution which is a combination of a Gaussian component and an
exponential function (for the disk),
\begin{eqnarray}
\label{decay}
\nonumber
\Sigma(\hmm) &=&	216\exp\left[-\left(\frac{r}{649 {\rm~pc}}\right)^2\right]+ \\
	& &	120\exp\left[-\frac{r}{2265 {\rm~pc}}\right]
		{\rm \ M_\odot \ pc^{-2}}
\end{eqnarray}
The Gaussian shape of the inner part of the distribution depends
mainly on the convolution of the true distribution with the beam (the
scale length 649 pc in Eq.\,\ref{decay} corresponds to a FWHM of
50\arcsec ).  The right panel of Fig.~\ref{expdec} shows also the mass
surface density of the HI gas.  These data were obtained from
\citet{TA93} (hereafter TA).  The column densities in TA were obtained
with the VLA, and the authors noted that their estimated total HI mass
within the Holmberg radius ($0.87\nine$) was lower than the
corresponding masses estimated by other authors using single dish
observations, e.g., both \citet{RLW74} and 
\citet{HB81} obtained about $1.6\nine$ (all masses recalculated to
the distance 4.5 Mpc).  The most probable explanation for the
deficiency was, according to TA, the absence of short-spacing
information.  In order to compensate for this we have added the
``missing'' mass surface density (2.4\,M$_\odot$\,pc$^{-2}$) to the HI
data of TA. The resulting HI mass surface densities are still low and
typically H$_2$ gas constitute about 70\% 
of the total mass of neutral hydrogen (H$_2$ + HI) in the disk inside 6.8\,kpc.

\begin{table*}
\caption{\COo\ and \COt\ velocity-integrated intensities of spectra
averaged within inclination-corrected annuli.  The annuli are fully
sampled out to 230\arcsec .  The \COt\ data were regridded to the same
spacing as the \COo\ so that the disk was sampled in the same way for
both transitions.  The difference in angular resolution in the data
sets has an effect on the response at the center.  The errors (1$\sigma$)
does not include the calibration uncertainty. }
\label{exptab}
\begin{tabular}{lccccc}
\hline
\hline
radius 
&  $I_{1-0}$& error
&  $I_{2-1}$& error
& \#spectra \\
{[}\arcsec ]
& [K\,km\,s$^{-1}$] & [K\,km\,s$^{-1}$]
& [K\,km\,s$^{-1}$] & [K\,km\,s$^{-1}$]
&in annuli\\
\hline
 	12 & 	 87.7 & 	 2.1 & 	 98.1 & 	 0.8 & 	 9 	\\
 	29 & 	 53.0 & 	 1.2 & 	 51.6 & 	 0.5 & 	 27 	\\
 	49 & 	 25.4 & 	 1.8 & 	 19.1 & 	 0.7 & 	 49 	\\
 	70 & 	 15.5 & 	 1.6 & 	 10.2 & 	 0.7 & 	 67 	\\
 	90 & 	 15.6 & 	 1.6 & 	 11.0 & 	 0.4 & 	 85 	\\
 	109 & 	 14.2 & 	 0.8 & 	 11.5 & 	 0.7 & 	 100 	\\
 	129 & 	 11.7 & 	 0.9 & 	 8.2 & 	 0.4 & 	 126 	\\
 	149 & 	 7.9 & 	 1.1 & 	 5.7 & 	 0.8 & 	 141 	\\
 	169 & 	 6.3 & 	 1.2 & 	 4.3 & 	 0.6 & 	 157 	\\
 	189 & 	 5.4 & 	 0.2 & 	 3.6 & 	 0.8 & 	 181 	\\
 	209 & 	 4.8 & 	 0.5 & 	 3.6 & 	 0.5 & 	 198 	\\
 	229 & 	 4.4 & 	 0.5 & 	 3.1 & 	 0.6 & 	 215 	\\
 	249 & 	 3.5 & 	 0.4 & 	 2.4 & 	 0.7 & 	 212 	\\
 	268 & 	 3.7 & 	 0.8 & 	 2.1 & 	 0.6 & 	 155 	\\
 	288 & 	 3.0 & 	 0.7 & 	 1.5 & 	 0.6 & 	 102 	\\
 	307 & 	 2.1 & 	 0.9 & 	 $<$2.7   & 	     & 	 44 	\\
\hline
\end{tabular}
\end{table*}

Based upon the H$_2$ mass surface density distribution (Eq.~\ref{decay})
we estimate a total molecular gas mass within a radius of 7.3 kpc
(the extent of the map) of $3.5\nine$, and extrapolated to the Holmberg
radius (7\farcm3 or 9.6 kpc) the mass becomes $3.9\nine$ (which agrees
well with the result of \citet{CTB02}, $3.7\nine$ (cf. also to the
result $3.3\nine$ that was derived in Sect.~\ref{molmass} using the
velocity-integrated intensity in the global spectrum).  
This is more than
twice the total HI mass within the same galactocentric radius.  It
should, however, be noted that 80\% of the HI mass in M\,83 is found
outside the Holmberg radius \citep{HB81}.
%
As a comparison, the stellar mass of M\,83 is $\sim$\,37$\nine$, 
given V$_T^0$=7.37 \citep{DDC92} and assuming a mass-to-light ratio of 2.

In the Gaussian/exponential disk decomposition the major part of the molecular
gas mass within the Holmberg radius lies in the disk, $3.6\nine$. 
\citet{L87} made \COo\ observations in 21 positions in the central parts
of M\,83, and he estimated the total molecular gas mass within
115\arcsec~to be 1.2$\nine$ (we have compensated for the difference
in \XCO~and distance).  Our H$_2$ distribution gives 1.4$\nine$.

\subsubsection{Comparison with data at other wavelengths}
\label{scalelenght}

\begin{figure}
	\resizebox{\hsize}{!}{\rotatebox{-90}
	{\includegraphics{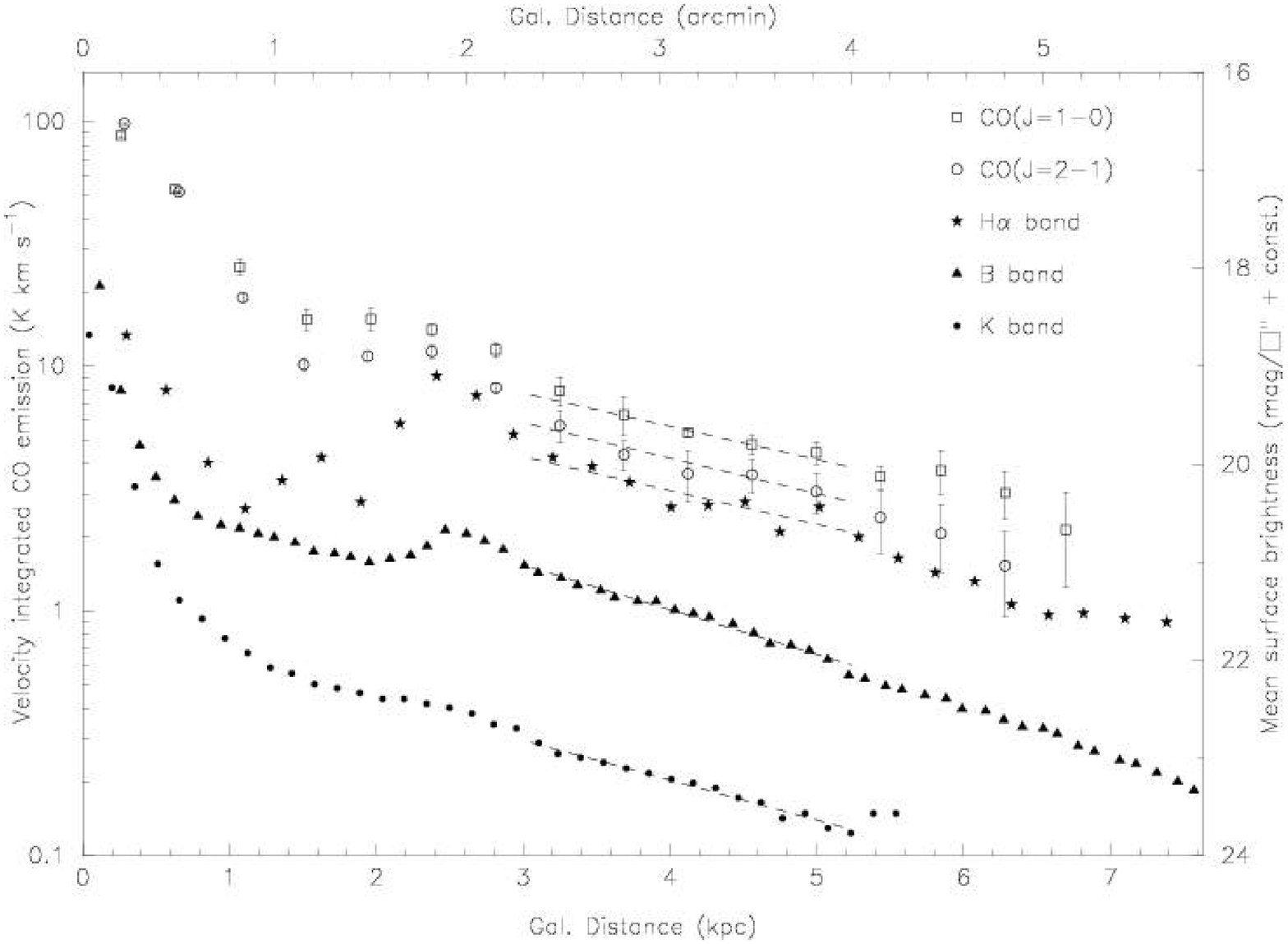}}}
	\caption{Azimuthally-averaged \COo\ and \COt\ velocity-integrated
	intensity (cf.~Table \ref{exptab}) in annular zones compared to surface
	brightnesses in the H$\alpha$, B, and K filters.
	Note how the bar ends appear as a bump at 120\arcsec~in the
	distributions.  The spatial resolution of the \COt\ is better than that of the
	\COo\ data, which results in the difference in slopes of the inner parts
	of these distributions. The dashed lines show the fits of 
	exponentials discussed in the text.
	\label{CObkha}}
\end{figure}

In Fig.~\ref{CObkha} we compare the CO radial distributions with
corresponding photometry data in the B, K, and H$\alpha$ bands.  The B and
H$\alpha$ data come from \citet{TJD79}, and the K data from
\citet{AAW87}.  The scale lengths of the various tracers compare well
over most of the radial range, but the comparatively low spatial
resolution affects the CO curves in the central region and at the bar
ends.  The latter are responsible for the peaks at $\sim$120\arcsec . 
We have fitted a single exponential to the data in the radial range
140\arcsec--240\arcsec, which corresponds to the region outside the
bar but inside the area where our CO survey is complete.  The scale
lengths are: 135$\pm$14\arcsec, 126$\pm$16\arcsec, 112$\pm$20\arcsec,
100$\pm$14\arcsec~and 132$\pm$35\arcsec~for \COo , \COt , K, B, and
H$\alpha$, respectively (the errors of the latter three do not include
the measurement errors).  Thus, the disk scale lengths are, within the
errors, the same for these star formation tracers.

The CO intensities decrease exponentially over the entire surveyed area,
and we do not observe the sharp cut-off at 4$\farcm$5 seen by
\citet{CTB02}.  In fact, we find CO emission with
$I_{1-0}$\,$\approx$\,1\,K\,km\,s$^{-1}$ in selected spectra as far from
the nucleus as 5$\farcm$5.  

\subsection{Brightness distributions}
We find that in the {\sc mem}-deconvolved map the average \COo\
intensity in the arms is about 2.5 times higher than in the interarm
regions.  Assuming a direct proportionality between CO emission and mass
surface density, this agrees well with the lower limit
of the arm--interarm mass surface density ratio, 2.3, found by RLH
using purely kinematical arguments.

\begin{figure*}
	\resizebox{0.95\hsize}{!}{\rotatebox{0}
	{\includegraphics{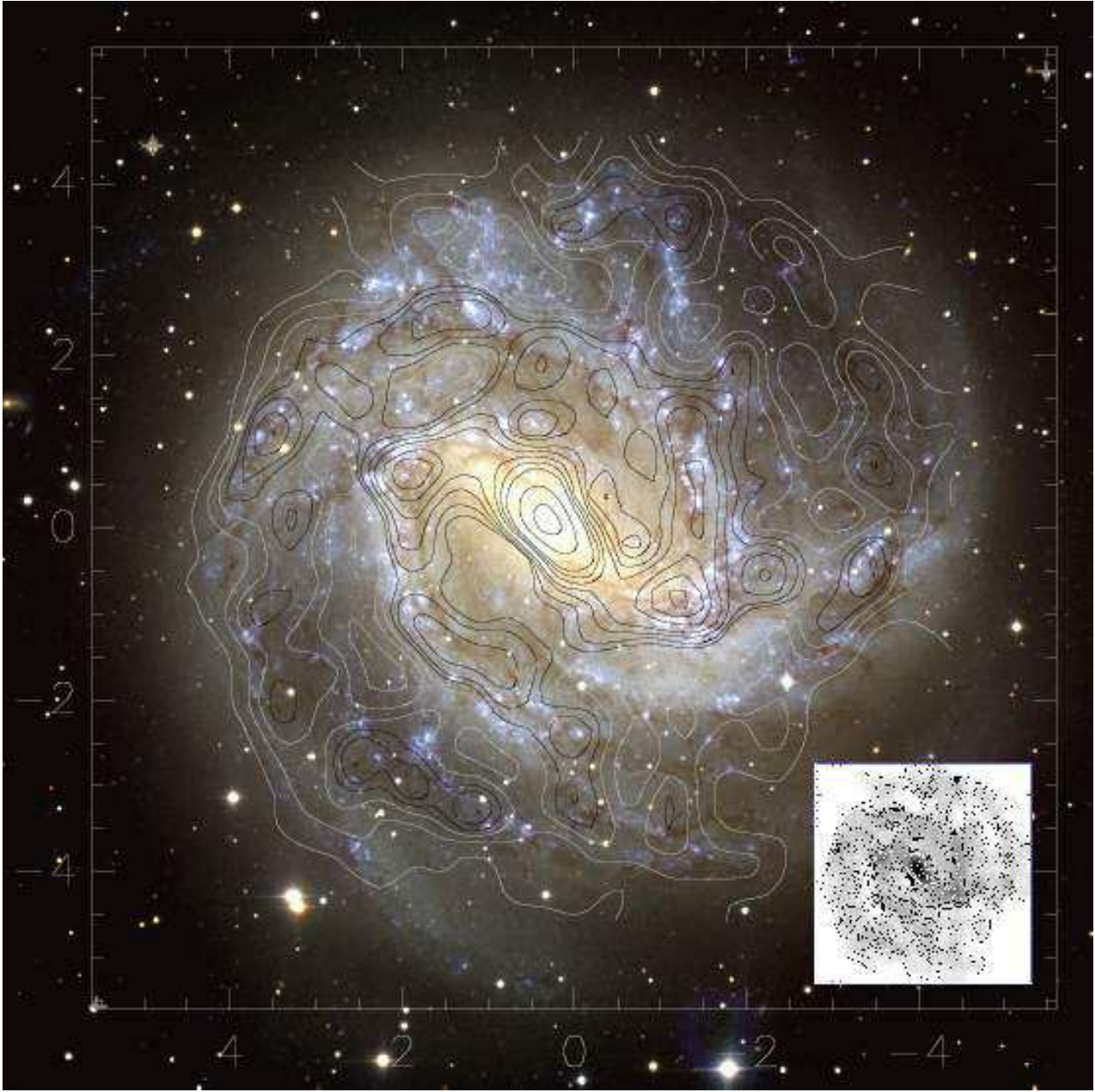}}}
	\caption{The {\sc mem}-deconvolved velocity-integrated \COo\ intensity
	as contours superposed on an RGB-map produced from images in B, V, and
	R. The optical images were obtained by S\"oren Larsen at the ESO
	Danish 1.5\,m telescope on La Silla, Chile.  Three stars have been
	used for image alignment.  These are located in the lower left corner
	and in both upper corners and are marked with crosses.  The inset
	shows the \COo\ map in gray-scale. }
	\label{COonRGB}
\end{figure*}

In Fig.~\ref{COonRGB} we show the {\sc mem}-deconvolved \COo\
velocity-integrated intensity map superposed on a true-color image. 
This image is a composite of three images in the filters B, V, and R,
rendered as blue, green, and red, respectively.  These images have
been obtained at the ESO Danish 1.54\,m telescope by S\"oren Larsen and the data
are described in \citet{LR99}.  On the large scale the \COo\ emission
follows the dust lanes, both in and between spiral arms, but at some
locations the \COo\ emission is displaced toward the regions where star
formation takes place \citep{WRH90}.  
Most notably, in the inner spiral arm on the
eastern side, the \COo\ ridge is located downstream of the dust lane
with a separation of about 10--15\arcsec\ (also shown in RLH), as are
the HI and H$\beta$ emissions as shown by TA. We note that in general
the OB associations straddle the rims of the \COo\ emission
concentrations, presumably due to extinction in combination with the
photodissociation by the young stellar population.  The outer and
inner spiral arms appear to connect at several locations, and these
connection regions often show an increased number of OB associations. 
The ``Gould belt structure'', at (1$\farcm$75,--2$\farcm$25) in the coordinate
system of Fig.~\ref{COonRGB}, is such a region \citep{C01}.

\begin{figure*}
	\resizebox{0.95\hsize}{!}{\rotatebox{0}
	{\includegraphics{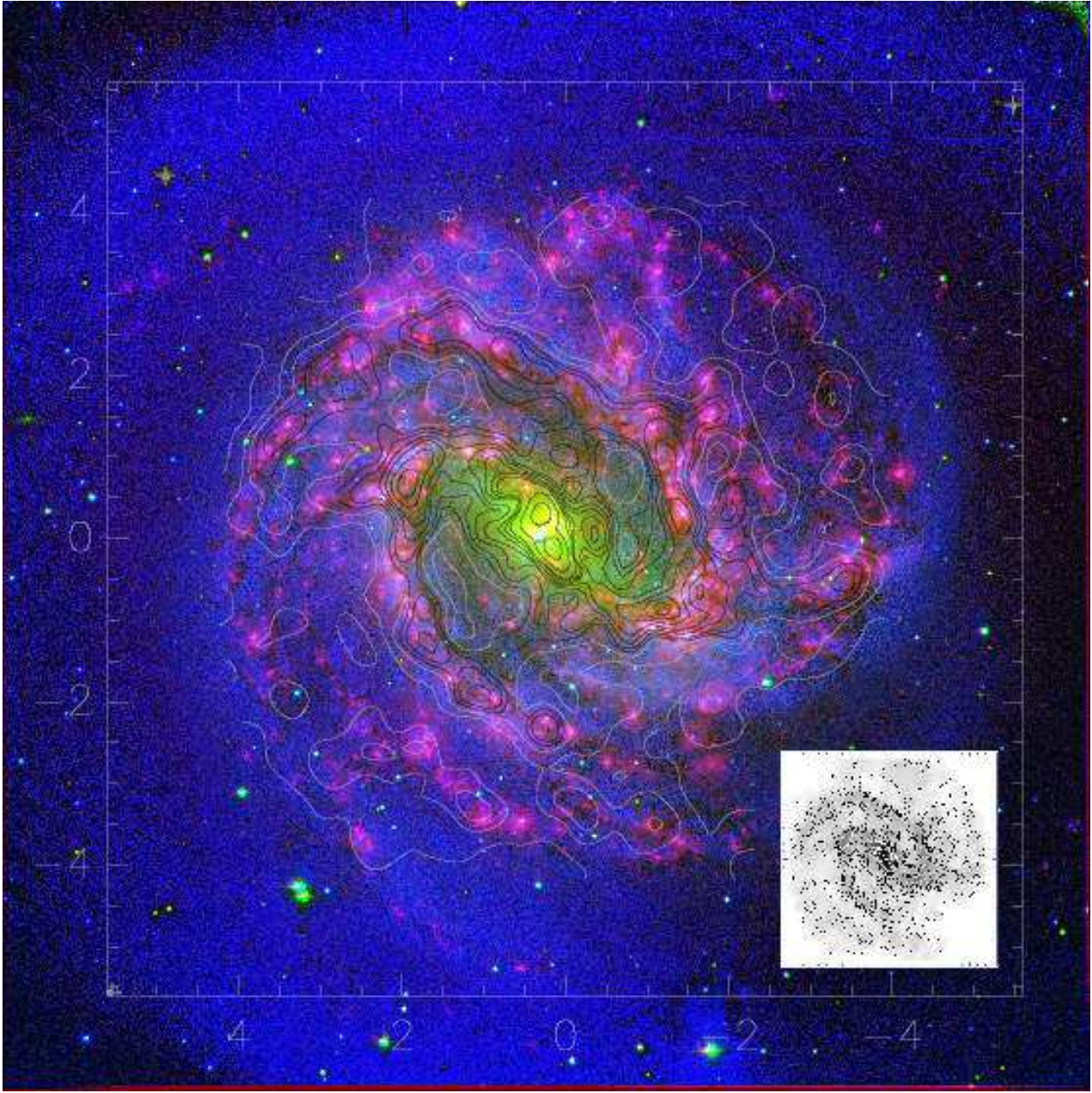}}}
	\caption{The {\sc mem}-deconvolved \COt\ velocity-integrated intensity as
	contours on an RGB image where the colors red, green and blue
	represent star formation (H$\alpha$), infrared emission (I band), and
	dust lanes (filters V--I), respectively.  The inset shows the \COt\
	map in gray-scale. }
	\label{CO21onRGB}
\end{figure*}

Figure~\ref{CO21onRGB} shows the \COt\ velocity-integrated intensity as
contours on an RGB image, where the colors represent star formation
(H$\alpha$; red), infrared emission (I band; green), and dust lanes
(filters V--I; blue).  In the disk, the observed \COt\ brightness
distribution strengthens the conclusions drawn from the \COo\ emission. 
In the central region the higher angular resolution at the \COt\
frequency reveals additional details.  There are two central nuclear
components relatively symmetrically placed on opposite sides of the IR
center.  At these locations the dust lanes, on the leading edges of
the bar, attach to the nuclear ring seen in J--K by \citet{ECW98}. 
Their color difference map (V--I) shows an area of high dust
extinction at the location of the NE component, whereas at the SW
component there appears to be relatively small amounts of dust.  It
appears as if the NE CO component lies above the disk (or the bulge) and
the SW CO component lies below or behind the same.  This was also
concluded by \citet{SW94} based on similar data.  Along
the bar, the \COt\ emission follows the dust lanes on the leading
edges.  In the disk the CO emission traces the H$\alpha$ emission
very well, although there are regions with a clear anti-correlation,
i.e., substantial CO emission but none, or very weak, H$\alpha$ emission,
such as the peak at about half an arcminute west of the nucleus.

\begin{figure*}
	\resizebox{\hsize}{!}{\rotatebox{0}
	{\includegraphics{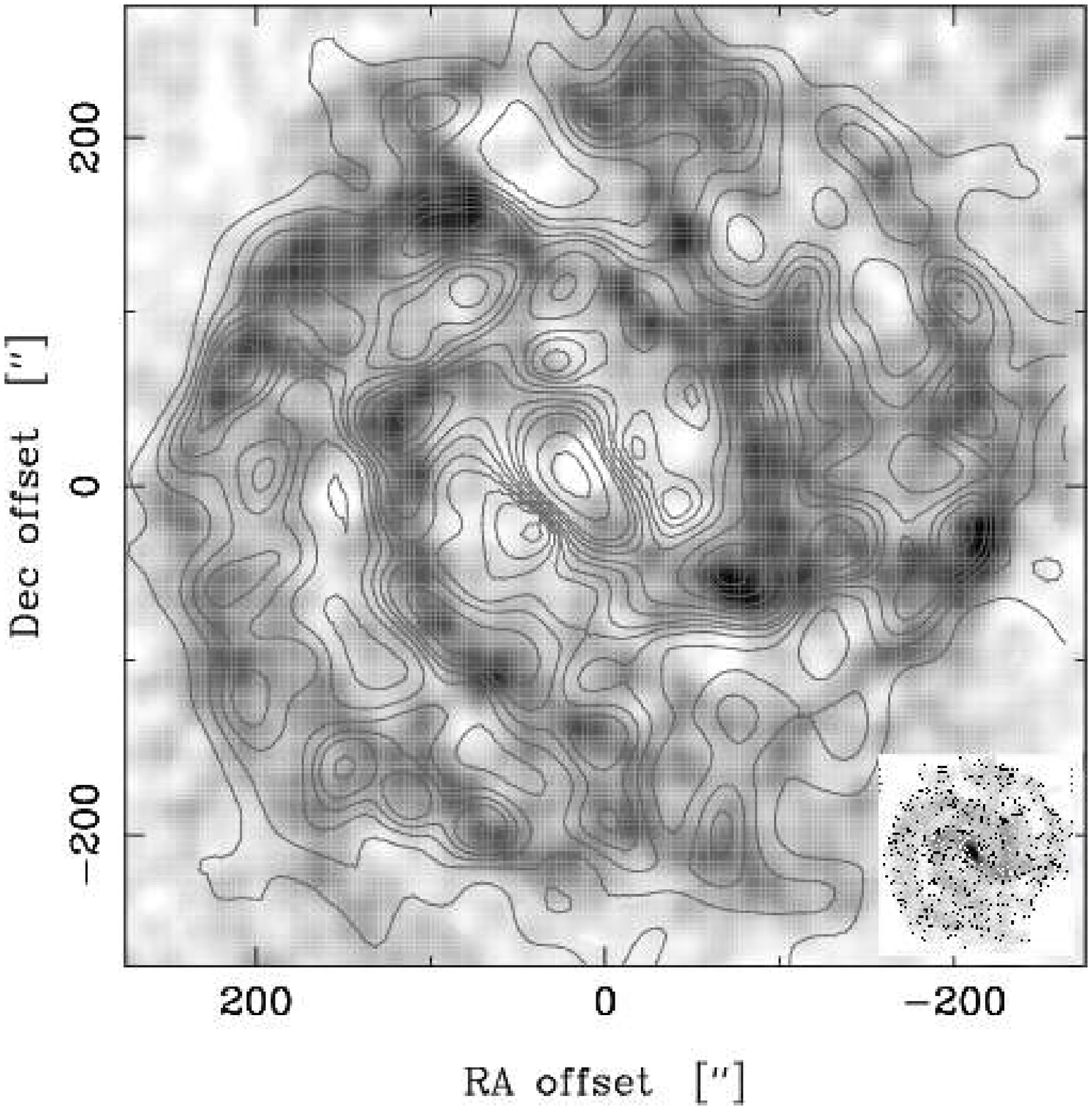}}}
	\caption{The {\sc mem}-deconvolved velocity-integrated \COo\ intensity
	as contours on an HI column density map (gray-scale) of roughly
	similar resolution ($\approx$20\arcsec ).  The HI data comes from TA.
	The inset shows the \COo\ map in gray-scale.}
	\label{CO10onHI}
\end{figure*}

Figure~\ref{CO10onHI} shows the \COo\ velocity-integrated intensity on a
gray-scale image of the HI column density (data kindly provided by 
R.P.J.~Tilanus and R.J.~Allen).  The angular
resolution of the HI map is slightly better than that of the CO map. 
The CO emission (which is assumed to be linearly proportional to the H$_2$
mass surface density intensity) and the HI column density  follow each other 
closely,
with one outstanding exception: the
nucleus, where very little HI is present.  At some places in the disk
the H$_2$ and HI maxima are displaced from each other, but there is no
apparent systematic trend that H$_2$ is outside HI or vice verse.
In general, the separations are of the order of 3-9$\arcsec$, which is small 
compared to the resolution in these maps (23$\arcsec$).

Local minima in the HI and H$_2$ gas surface densities also correlate
well.  At (50$\arcsec$,180$\arcsec$) in the coordinate system of
Fig.~\ref{CO10onHI}, marked minima can be seen in both H$_2$ and HI.
The total gas surface density in this area is about
3--6\,M$_\odot$\,pc$^{-2}$, of which only
$\approx$1\,M$_\odot$\,pc$^{-2}$ is in the form of molecular gas. 
Such low mass surface densities should not give rise to any massive
star formation, and indeed, the H$\alpha$ image shows very little
emission here.

\begin{figure*}
	\resizebox{\hsize}{!}{\rotatebox{0}
	{\includegraphics{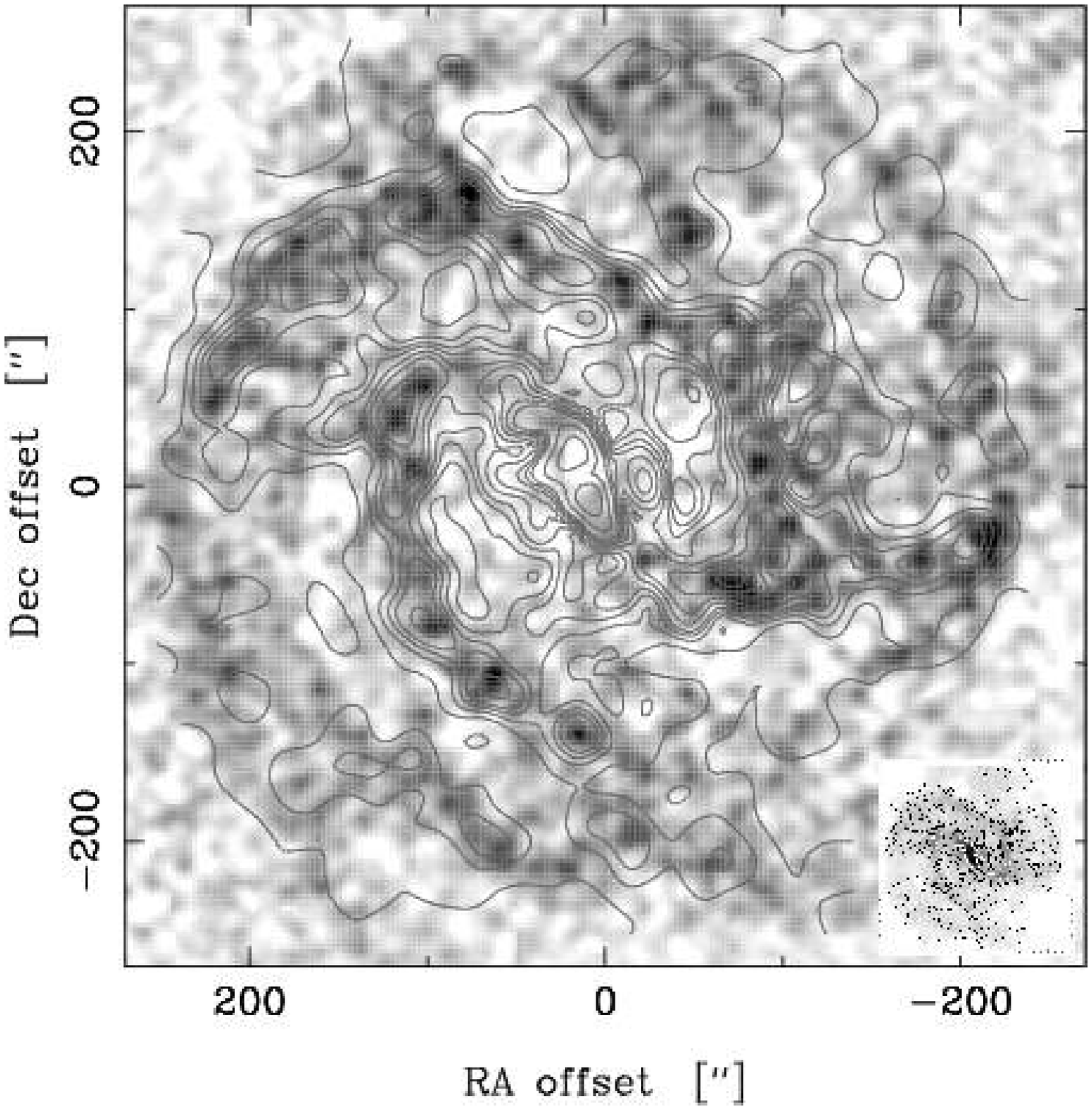}}}
	\caption{The {\sc mem}-deconvolved velocity-integrated \COt\ intensity
	as contours on an HI column density map (gray-scale) of roughly
	similar resolution ($\approx$12\arcsec ).  The HI data comes from TA.
	The inset shows the \COt\ map in gray-scale.}
	\label{CO21onHI}
\end{figure*}

Figure~\ref{CO21onHI} compares the H$_2$ (as probed by the \COt\
emission) and the HI gas in more detail, the resolution being
$\approx$\,10$\arcsec$.  In the main disk the \COt\ emission and HI
emission follow each other closely and several of the concentrations
in the \COt\ emission are also seen in the HI distribution.  Again the
exception is the nuclear region.  The mass surface density of HI at
the two nuclear components differ markedly: the SW component has a
peak mass surface density of 6\,M$_\odot$\,pc$^{-2}$, while at the
center of the NE component HI is actually seen in absorption
against a nuclear
continuum source, which again speaks in favor of the idea that the SW
and NE components lie on different sides of the disk (or a bulge). 
The mass surface density of HI at the center is very much lower
than that of H$_{2}$, where the highest value in the $I_{1-0}$ map
corresponds to $\approx$750\,M$_\odot$\,pc$^{-2}$.  In the $I_{1-0}$
map it is not possible to distinguish the separate nuclei due to the
relatively low spatial resolution and the coarse grid.  However, using
the average \mbox{2--1/1--0} line ratio (see Sect.~\ref{secratio}), the mass surface density is $\sim$1100\,M$_\odot$\,pc$^{-2}$ for each of these components. 
As the physical properties of the molecular gas in the center may
differ from that in the disk, application of the standard conversion
factor may lead to an erroneous result.
Indeed, it is likely that at least part of
the emission in the center comes from a diffuse and
non-virialized component. This would lead to an 
overestimate of the molecular gas mass \citep{PA00}.

The explanation for the nice correlation may be that the HI 
is mostly a dissociation product, as TA discuss.  A smooth distribution of
`primordial' HI gas would not be detected in the interferometer
observations (in which roughly 50\% of the HI gas was detected) while
HI produced as an effect of dissociation of molecular gas in the
neighborhood of star formation would be localized, and therefore
detected by the interferometer.


\subsection{The CO ($J$=2--1)/($J$=1--0) line intensity ratio}
\label{secratio}

\begin{figure}
	\resizebox{\hsize}{!}{\rotatebox{-90}
	{\includegraphics{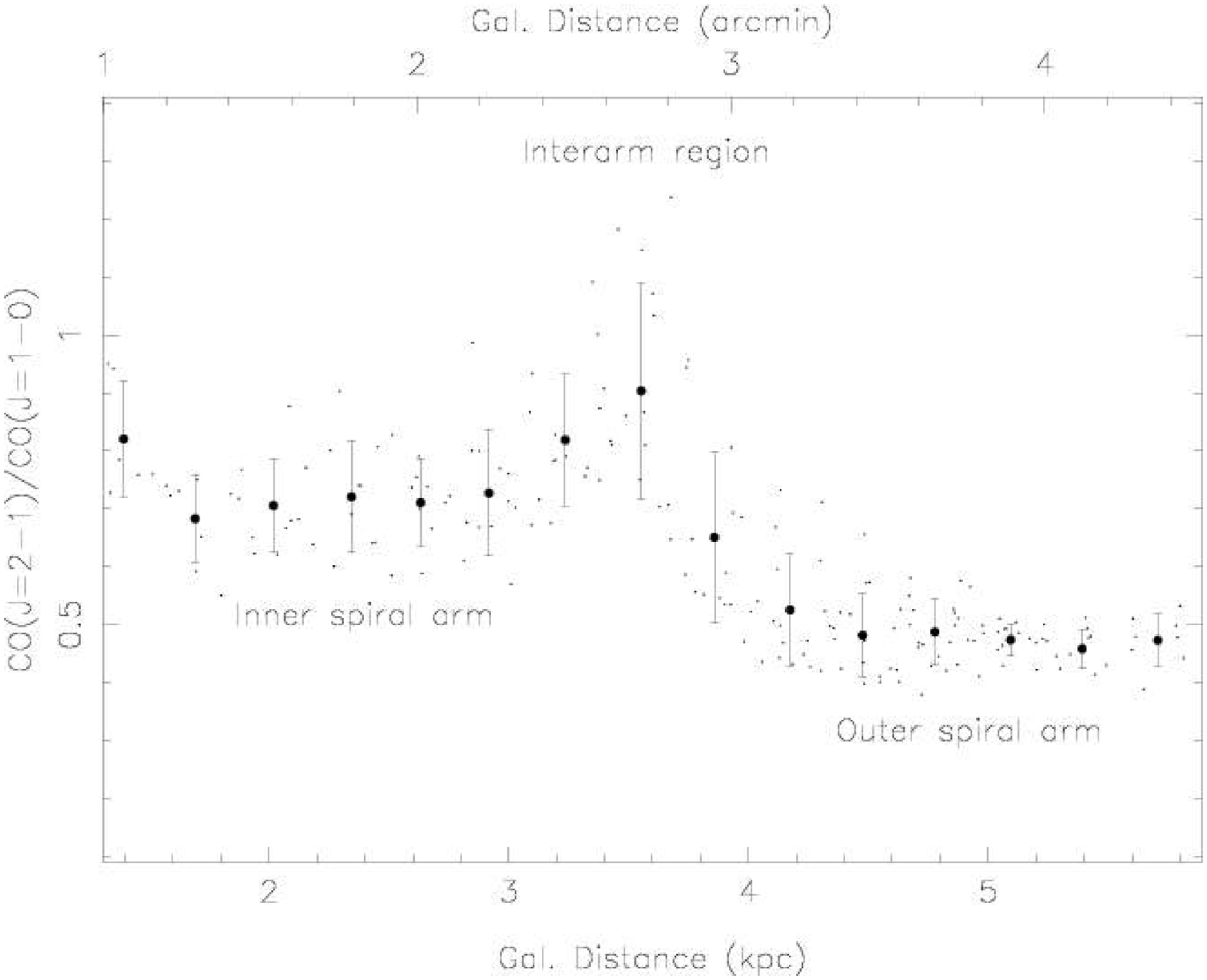}}}
	\caption{The CO ($J$=2--1)/($J$=1--0) line intensity ratio, $R_{21}$, as
	a function of galactocentric radius. The interarm region shows a
	clear trend of increased line ratio. The data are taken from a 
	$45\degr$ wide sector SE of the nucleus (position angle 90\degr --135\degr).}
	\label{ratio_sec}
\end{figure}

We have created a map of the CO($J$=2--1)/($J$=1--0) line intensity 
ratio, $R_{21}$, at 49\arcsec~resolution. The mean value over this 
map is 0.72$\pm$0.19, which is within the errors the value derived
from the velocity-integrated intensities in Table \ref{comp12}. 
\citet{CTB02} estimate a ratio of 1.1$\pm$0.2, which we suspect is
higher than ours because they have not corrected their \COt\ data for the
error beam contribution (considering the main beam efficiency of the
NRAO 12\,m telescope at this frequency, $\approx$0.56, the error beam is likely
to be comparable to that of the SEST).  In the nuclear region and along the
bar the ratio is 0.83$\pm$0.04. 
The $R_{21}$ map shows also that on the arms,
the line intensity ratio is lower than in the interarm regions. This is 
illustrated in Figure \ref{ratio_sec}, where $R_{21}$ is plotted as a function
of radius in a $45\degr$ wide sector of the galaxy (position angle 
90\degr --135\degr ). In order to quantify and compare the ratio on the arms
to that in the interarm regions, we have selected two sectors where the
arms are well separated. The position angle of these sectors are: 
90\degr --135\degr~(SE) and 270\degr --315\degr~(NW). 
In these sectors we selected spectra belonging to the inner, outer and 
inter-arm regions, and averaged them separately. Finally,
``arm''-spectra were created by taking the average of the inner and outer arm
spectra in the two respective sectors. From these eight spectra, four
ratios were calculated. 
Assuming our error beam is correct (Sect. \ref{apperrbeam}),
the line ratio is significantly lower on
the arms: 0.59$\pm$0.04 (SE) and 0.69$\pm$0.04 (NW) compared to the
ratio in the interarm region 0.88$\pm$0.17 (SE) and 0.98$\pm$0.21 (NW) 
(the errors are 3$\sigma$ and are estimated from the noise level in the spectra).
We believe that this is an effect of a (partly) different cloud component in
the interarm regions, where the heating of the gas is more efficient
than in the dense arm cloud complexes. 
As a note: We have also calculated the
$R_{21}$ value for the {\sc mem}-deconvolved \COo\ and \COt\ maps at
the common resolution 30\arcsec .  The arm value is roughly the same
(0.6$\pm$0.1), but the interarm value is higher, 1.3$\pm$0.3. 

\begin{figure}
	\resizebox{\hsize}{!}{\rotatebox{-90}
	{\includegraphics{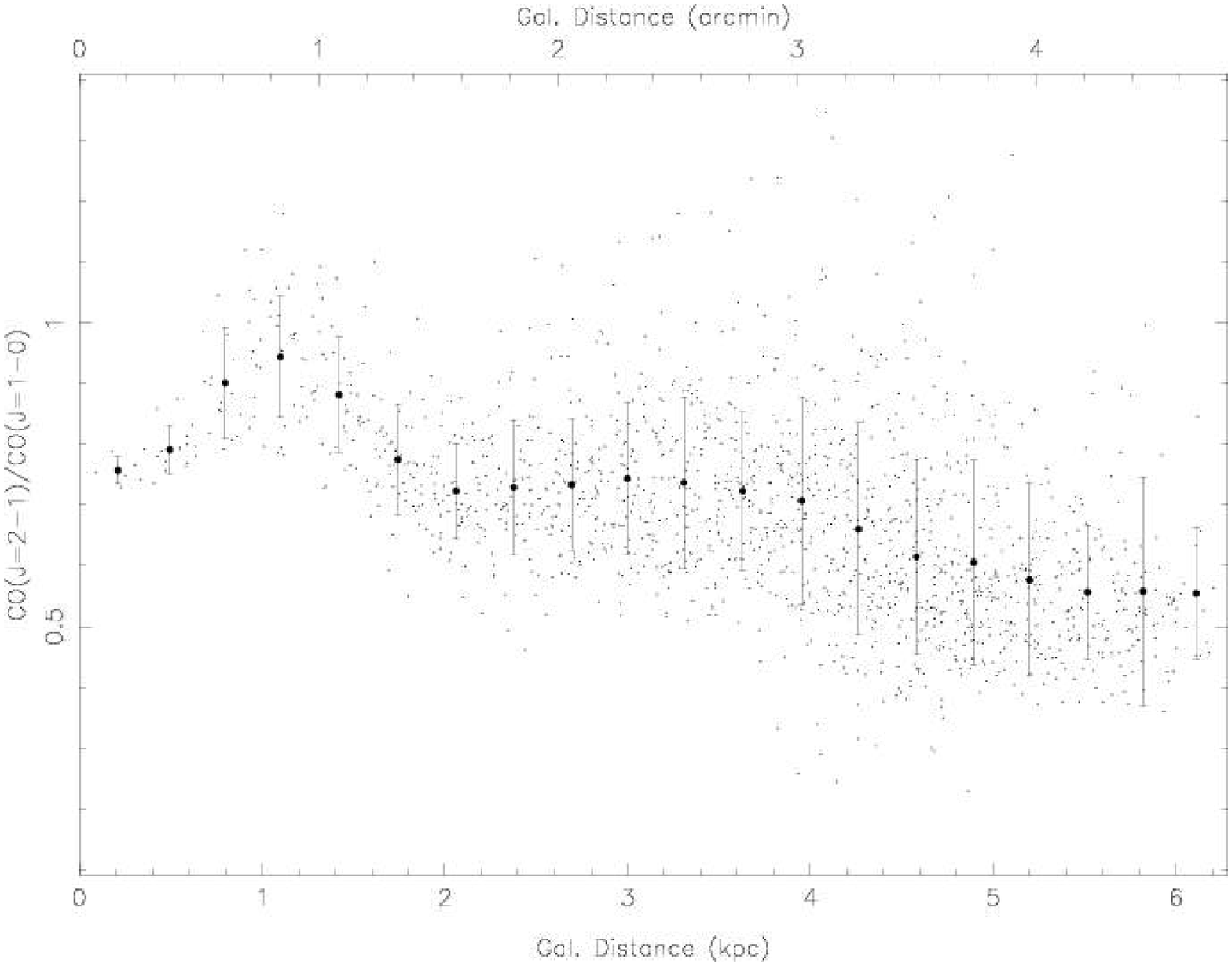}}}
	\caption{The CO ($J$=2--1)/($J$=1--0) line intensity ratio for individual
	positions (dots) and the azimuthally averaged ratio (with error bars)
	in annular zones in the plane of the galaxy. }
	\label{radratio}
\end{figure}

By averaging $R_{21}$ in concentric annuli, we find a trend in the line
intensity ratio in the sense that it decreases with galactocentric
distance, Fig.~\ref{radratio} (this is also indicated in
Fig.~\ref{CObkha}).
In particular, the lower envelope of the $R_{21}$-values is clearly decreasing. 
A similar difference between the $R_{21}$ line ratio in the disk and towards
the nucleus has previously been shown to exist in our Galaxy \citep{SHH01}.
The wiggles in Fig.~\ref{radratio} can be
associated with features in the $R_{21}$ map. The nucleus appears to be
surrounded by a ring of excited gas, which is reflected as a bump in
the radial distribution.  The bar ends and inner arms are reflected as a
flat region at $\approx$120--150$\arcsec$ , while the low
$R_{21}$-values in the outer arms suppresses the average
$R_{21}$-ratio in the range 200--300\arcsec .

\section{Conclusions} We have mapped, in over-sampling mode, the \COo\ and \COt\ brightness
distributions over the entire optical disk of M83.  Significant corrections
for an error beam have been applied to the \COt\ data.  
A {\sc mem}-method has been used to increase the angular
resolution of the data.  

We find that:

{\it i)} the CO emission is strongly peaked toward the nucleus, which splits
up into two components in the \COt\ data. Also the bar ends are
prominent.  The CO emission follows the leading edges of the bar.

{\it ii)} Molecular gas spiral arms are clearly identified, and they trace,
in most cases, the dust lanes.  There are frequent ``bridges'' between
spiral arms, and in some areas the arms are clearly disturbed.

{\it iii)} An average arm--interarm brightness contrast of about 2.5 is found
for the \COo\ line.  

{\it iv)} The CO (\mbox{$J$=2--1})/(\mbox{$J$=1--0}) line ratio, which is
about 0.77 on average, differ between the arm and interarm regions. 
It is significantly higher in the latter. 

{\it v)} Regularly spaced molecular gas concentrations of mass
$\approx$10$^7$\,M$_{\odot}$ lie along the arms.

{\it vi)} The estimated total molecular gas mass is
3.9$\times$10$^9$\,M$_\odot$, and within 7.3\,kpc the H$_2$ mass
dominates over that of HI by a factor of more than two.  The estimated
gas/stellar mass ratio is $\approx$0.1 in the optical disk.

{\it vii)} The CO and HI emissions are very well correlated in the
optical disk, and frequently concentrations of the two gas components coincide.

{\it viii)} The CO radial brightness distribution in the disk follows
that of other starformation tracers as H$\alpha$ emission, and
continuum light in the B and K filters.  The estimated scale length,
of a fitted exponential, is about 120$\arcsec$, corresponding to
2.6\,kpc at the adopted distance.

\begin{acknowledgements}
We are very grateful toward the SEST staff for their support
during observations and Swedish Natural Science Research Council
for travel expenses support.
We also wish to thank Remo Tilanus and Ron Allen
for letting us use their HI data and S\"oren Larsen for the use of the
optical images and the referee for valuable input. 
\end{acknowledgements}

\appendix
\section{Error beam} \label{apperrbeam}
The beam pattern of a radio telescope is affected by
its surface accuracy. This means that the power received in the 
error beam may very well be comparable to the power received in the
main beam. For sources that are compact relative to the extent of the
beam, it is relatively easy to compensate for this underestimate
of the true flux: The measured antenna temperature ($T_{\rm A}^{*}$) is
simply divided by the main beam efficiency, $\eta_{\rm mb}$, to attain the
main beam brightness temperature ($T_{\rm mb}$). The main beam efficiency
is usually well known and can be found in the user manual of the different
telescopes. For extended sources, the situation is more complex, since 
the antenna may receive power not only from the main beam. If the size of
the error beam is comparable to the extent of the source, a
simple multiplication of $(\eta_{\rm mb})^{-1}$ will overestimate the
flux at the position where the measurement is done. If the antenna pattern
is well known, it is possible to calculate how the beam couples to the
source and thus compensate for this effect. Unfortunately, the antenna
pattern for SEST has never been measured properly, so this was not an
option for us. Instead we devised a procedure to remove the impact of 
the error beam, using our own data set to estimate the strength and size
of the error beam pattern.

Nearly all \COt\ spectra did show two components: one narrow line, that agrees
with the width and center velocity of the \COo\ spectra at the same position
and another, broader, component that is picked up by the error beam.
By studying these broad wings in a number of spectra in the deconvolved
\COt\ data set and by comparing with spectra
in data cubes convolved with different FWHPs, we estimated that the size
of the error beam at 230 GHz is of the order of 200\arcsec .  This is
in good agreement with the error beam size of 3\arcmin, which was assumed by
\citet{JGB98}.

In order to try to estimate the impact of the error beam we produced a data
set where we convolved the {\sc mem}-deconvolved data set with a 
200\arcsec~Gaussian profile.  In each position an automated process, using the
sliding window technique, calculated the velocity-integrated intensity in
the region of the extended wings [i.e. outside the velocity range
where we have seen emission in the \COo\ data] in both the deconvolved
cube and in the 200\arcsec~convolved cube.  The ratio between these
intensities corresponds to the fraction of the available emission which is
picked up by the error beam.  We will refer to this as the error beam
efficiency.  
In principle, the error beam
efficiency must lie in the interval 0.1--0.5, since the main beam
efficiency is 0.5 and the moon beam efficiency is 0.9.  Some values of our
error beam efficiency estimates are higher than 0.5.  This
happens in regions where the emission in the main beam falls outside
the expected velocity range (e.g., in regions with exceptionally
strong streaming motions or wide emission profiles), leading to an
over-estimation of the flux in the wings.  An attempt to make the
window wider had a negative impact on the results since we were left
with fewer channels to calculate the velocity-integrated emission in
the wings.  Neglecting values above 0.5, we found that the error beam
efficiency was on average 0.27$\pm$0.11.

\begin{figure}
	\resizebox{\hsize}{!}{\rotatebox{-90}
	{\includegraphics{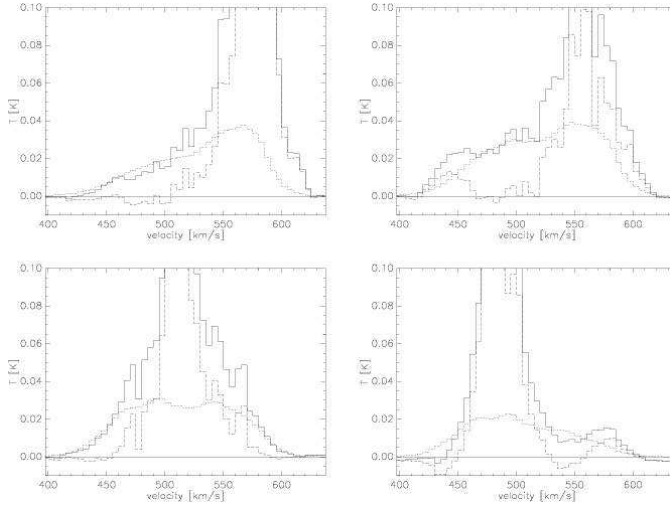}}}
	\caption{These spectra show the original spectra (solid lines),
	200\arcsec~convolved spectra (short-dashed lines) and
	subtracted spectra (long-dashed lines) in four typical positions.
	The coordinates with respect to the map center is, from upper
	left to lower right, (-42,-70), (-42,0), (-42,70) and (-42,140).
	The 200\arcsec~convolved spectra are scaled with the derived error 
	beam efficiency.
	}
	\label{errbspec}
\end{figure}

Figure \ref{errbspec} shows the original spectrum, the 200\arcsec~convolved
spectrum (scaled with the error beam efficiency), and the resulting (cleaned)
spectrum after subtraction of the two previous spectra, in four
typical positions.  The convolved spectra reproduces the shape of the wings
very well, and the shape of the cleaned spectra resembles the
corresponding \COo\ spectra (not shown).

We have corrected the convolved and the {\sc mem}-deconvolved data sets
for this error by subtracting the error beam contribution for each
individual spectrum.  Based on the previous result, we took a
conservative approach and fixed the error beam contribution to 0.27. 
In principle, we could have let this ratio vary from spectrum to
spectrum by fitting the 200\arcsec -convolved spectrum to the extended
wings, but this would have introduced other, not easily controlled,
errors.

\bibliography{H4219BIB}
\bibliographystyle{aa}

\end{document}